\documentclass[useAMS,usenatbib,a4paper]{mn2e}
\usepackage{times}
\title[kpc regions of the ISM]
{
The Interstellar Medium and star formation on kpc size scales
}
\author[Dobbs]
{Clare L. Dobbs\thanks{E-mail:
dobbs@astro.ex.ac.uk}$^{1}$ \\
$^1$ School of Physics and Astronomy, University of Exeter, Stocker Road, Exeter, EX4 4QL, UK \\
}
\usepackage{amssymb}
\usepackage[pdftex]{graphicx}
\usepackage{epsfig}
\usepackage{multirow}

\begin{document}
\label{firstpage}
\date{\today}

\pagerange{\pageref{firstpage}--\pageref{lastpage}} \pubyear{2012}

\maketitle

\begin{abstract}
By resimulating a region of a global disc simulation at higher resolution, we resolve and study the properties of molecular clouds with a range of masses from a few 100's~M$_{\odot}$ to $10^6$~M$_{\odot}$. The purpose of our paper is twofold, i) to compare the ISM and GMCs at much higher resolution compared to previous global simulations, and ii) to investigate smaller clouds and characteristics such as the internal properties of GMCs which cannot be resolved in galactic simulations.
We confirm the robustness of cloud properties seen in previous galactic simulations, and that these properties extend to lower mass clouds, though we caution that velocity dispersions may not be measured correctly in poorly resolved clouds. We find that the properties of the clouds and ISM are only weakly dependent on the details of local stellar feedback, although stellar feedback is important to produce realistic star formation rates and agreement with the Schmidt-Kennicutt relation. We study internal properties of GMCs resolved by $10^4-10^5$ particles. The clouds are highly structured, but we find clouds have a velocity dispersion radius relationship which overall agrees with the Larson relation. The GMCs show evidence of multiple episodes of star formation, with holes corresponding to previous feedback events and dense regions likely to imminently form stars.  Our simulations show clearly long filaments, which are seen predominantly in the inter-arm regions, and shells.
\end{abstract}
\begin{keywords}
galaxies: ISM, ISM: clouds, ISM: evolution, stars: formation
\end{keywords}

\section{Introduction}
The scales between molecular clouds and galaxies are not well understood, and difficult to study either observationally or numerically. 
Whilst we are able to resolve intricate structure of the ISM in the Milky Way, our position within the Galaxy means it is very difficult to map the ISM on kpc scales. Conversely in other galaxies we can view the disc face on, but do not have the resolution to study scales below a GMC \citep{Colombo2014}. In simulations we are limited by the resolution required to simultaneously model star forming scales and larger scale processes.

Although difficult to map in Cartesian coordinates, the nearby Milky Way does provide a wealth of observations on features of size scales lower than GMCs. The number of GMCs that we are able to study in the Milky Way is small, limited to a few complexes such as Orion, Carina California, and Sco-Cen. In contrast the vast majority of clouds we are observe are only $\sim10^4$ M$_{\odot}$ \citep{Heyer2009}. Understanding the properties and evolution of such clouds requires exploring the scales between individual star formation and GMCs. Similarly we can observe features such as shells and interstellar filaments on scales of 10's pcs, again lying between the scales of star formation and galaxies typically adopted for simulations. These appear to be ubiquitous across the ISM, but recent observations have highlighted the existence of extremely long, $>100$ pc thin filaments along spiral arms and inter-arm regions \citep{Li2013,Ragan2014,Wang2014,Goodman2014}.

With regards numerical simulations, many works have considered isolated galaxies \citep{Wada1999,Wada2000,Wada2008,Dobbs2008,Tasker2009,Dobbs2011new,Dobbs2013,Renaud2013}. However these simulations can typically only resolve GMCs ($>10^5$ M$_{\odot}$), on scales of $\sim10$ pc, and do not provide any information on smaller clouds, and do not resolve well shells and filaments. Some simulations have focused on studying stellar feedback, in particular identifying the nature of the vertical structure and what drives winds and outflows from the disc \citep{Rosen1995,deAvillez2000,deAvillez2004,Slyz2005,Kim2011,Kim2013,Hill2012}. However such simulations only model a very small region of the galactic disc, and do not capture large scale processes such as spiral shocks and larger scale gravitational instabilities. \citet{Smith2014} do perform global disc simulations which are able to reach high resolution in one area of the disc, but these do not include self gravity or stellar feedback which significantly alters the structure of the ISM.

One approach used recently to incorporate galactic scale processes and resolve the detailed structure and kinematics of clouds is to resimulate regions of global simulations at higher resolution \citep{Bonnell2013,vanLoo2013}. This has enabled the study of GMCs at much higher resolution, to consider their internal properties and star formation rates. In these simulations (which use SPH and AMR respectively), star formation appears to be correlated roughly according to the Schmidt-Kennicutt relation \citep{Schmidt1959,Kennicutt1989}, but at values that are much too high. However in neither case do either the original global calculations, or resimulations include stellar feedback, and in the case of \citet{Bonnell2013} the original global simulations did not include self gravity. This led both papers to suggest that stellar feedback, and perhaps magnetic fields would be required to produce lower star formation rates. The simulations also appear to show that the clouds have substantial internal velocity dispersions which means that they are either unbound or only marginally bound. 

In this paper, we resimulate a region from the simulation shown in \citet{Dobbs2013}, which modelled a galactic disc with an $N=2$ spiral potential, stellar feedback, self gravity and heating and cooling.  This simulation was able to well resolve GMCs of mass $10^5$ M$_{\odot}$ but struggled to resolve GMCs of $10^4$ M$_{\odot}$. In the results presented here, we are able to resolve and study the properties of clouds down to masses $<1000$ M$_{\odot}$. After describing the details of our simulations (Section~2), we divide our results into three sections. These cover the structure of the ISM (Section~3), properties of clouds (Section~4) and star formation rates (Section~5). Within each section we compare our resimulaions with the original global simulation \citep{Dobbs2013}, and compare different feedback schemes. We also examine structures or properties that we could not resolve well in the global simulations, such as shells and filaments, the internal properties of GMCs and the Kennicutt Schmidt relation on sub galactic scales.

\section{Details of simulations}
In this paper we resimulate a section of the galaxy-scale simulation presented in \citet{Dobbs2013}, modelled using Smoothed Particle Hydrodynamics (SPH), at higher resolution. We select a region at a time frame of 250 Myr of size 1 kpc by 1 kpc along a spiral arm. Then we trace the gas particles in that region back by 50 Myr to a time of 200 Myr (similar to the procedure in \citealt{Bonnell2013}). At 200 Myr, this gas is situated in an elongated feature, primarily lying between two spiral arms. We further include any particle from the original galaxy simulation that is located within the area occupied by the boundary of the outermost particles of this region. We do not introduce boundary conditions, or different size particles (which is not usually recommended in SPH) around the edges of our region. Instead to our set of particles, we add any particles which lie within a certain distance of those already selected.
We choose this distance to be the minimum of the smoothing length of each particle and 70 pc (particles rarely have such large smoothing lengths unless they lie a long way above or below the plane). These neighbouring particles add another $\sim$20,000 particles to our original set. This leaves us with 80,330 particles.
We then split each of these particles into 81 particles, after which, we have a total particle number of 6,506,730. As the mass of the particles in \citet{Dobbs2013} is 312.5 M$_{\odot}$, the mass of the each particle in our resimulation is then $\sim$3.85 M$_{\odot}$.

All our calculations include self gravity, heating and cooling, and simple H$_2$ and CO formation as described in \citet{Dobbs2008}, and \citet{Pettitt2014}. The minimum temperature of gas in the simulation is 50 K. We also include a 2 armed spiral potential, as was included in the original simulation from \citet{Dobbs2013}. 
We choose a density threshold of 500 cm$^{-3}$ to insert feedback. As well as lying above the density threshold, gas also has to be bound and a converging flow (i.e $\nabla \cdot \mathbf{v}<0$) for feedback to occur.

We carry out simulations with a number of different feedback schemes, and overall perform 5 simulations. The prescription we use nominally represents supernovae feedback, and includes thermal and kinetic feedback. However as we typically insert energy immediately, or for some duration after star formation has assumed to occur, our feedback may in some ways better represent stellar winds which deposit a similar amount of energy into the ISM \citep{Agertz2013}. As described in Dobbs et al. 2011, we determine the  mass of molecular gas from a region of particles (typically 30 to 50) and multiply by an efficiency $\epsilon$ to find the amount of star forming gas. The amount of energy inserted is
\begin{equation}
E=\frac{\epsilon \times M(H_2) \times 10^{51}}{160 M_{\odot}}  ergs
\end{equation}
where $M(H_2)$ is the mass of molecular hydrogen and we divide by 160 M$_{\odot}$ (a value determined by taking a Salpeter IMF) to obtain the number of massive stars, and $10^{51}$ ergs is the energy of one supernova. The energy is inserted into the same particles used to calculate M(H$_2$), and is distributed evenly between the particles. Two thirds of the energy is deposited as kinetic energy, and one third is deposited as thermal energy. We do not expect overcooling to have a large effect as tests of the global simulations indicated that inserting energy solely as thermal or kinetic energy resulted in relatively small differences in the gas morphology and temperature distribution, and we have higher resolution here. 

Firstly we include feedback using the same method as described in \citet{Dobbs2011new}, where feedback is instantaneous. We are now including feedback on a much smaller scale ($\sim$ pc), so we choose higher efficiency parameters compared to \citet{Dobbs2013}. We performed simulations with efficiencies of $\epsilon=$ 0.1, 0.25 and 0.4. By comparison, in \citet{Dobbs2013} we chose $\epsilon=$ 0.05. For the simulations here, $\epsilon=$ 0.4 and $\epsilon=$ 0.1 represent relatively high and low degrees of feedback, and have correspondingly large and small effects on the disc.
The $\epsilon=$ 0.25 case was very similar to that with $\epsilon=$0.4 so we don't show any results from this simulation in the paper. A summary of the simulations ran is shown in Table~1, except for the $\epsilon=$ 0.25 case which is not discussed further.

In all of the simulations described so far, our resolution is so high that the number of massive stars ($\epsilon M(H_2)/160 M_{\odot}$) is less than 1. In our 3rd simulation (Run 3), we apply a stochastic approach so that (by generating random numbers) only 10 per cent of regions that satisfy our critical density criterion are assumed to form stars. But for those cases, the efficiency factor is multiplied by a factor of 10, which equates to the formation of $\sim$1 massive star (here we set specifically that each star formation event corresponds to one massive star). In practice this is equivalent to saying that only occasionally does a $\sim 100$ M$_{\odot}$ region form a massive star, in our simulation, this is about 10 per cent of the time.

In practice, our stochastic method proved to be not very different from Runs 1 and 2. This is because the code checks whether to implement feedback during each timestep, and the time steps are  quite small ($\sim1000$ years). Hence a region will statistically be likely to undergo feedback within ten timesteps, which is not very different from instantaneous feedback. We tried using large efficiencies, however this requires inserting larger amounts of energy which proved problematic and required prohibitively small time steps. Ideally, we would subtract some amount of mass and convert it to stars, whether or not a massive star is formed, but we do not consider this level of detail. With a relatively low efficiency the total number of star formation events is not so dissimilar from Runs 1 and 2, so effectively the result is a feedback scheme more similar to Runs 1 and 2 with an efficiency of 0.25.

We ran three calculations with the stochastic prescription, one with the instantaneous feedback description described above. For the others (Run~4 and 5), stellar feedback is inserted over time. We choose relatively short periods, 2 and 5 Myr, because the timescales of the simulations are quite short. 
 With calculations where the feedback is inserted over time, the feedback tends to be less effective, hence a higher star formation efficiency is used compared with Run 3. Also with feedback being continuously inserted, there may be overall fewer events compared to Run 3. This means that in Runs~4 and 5, each feedback event corresponds to roughly one massive star forming (see also next paragraph).
 
Runs~4 and 5 also include star particles. One gas particle is changed to a star particle per feedback event (the same as described in \citealt{Dobbs2014}). The star particles only experience gravity, but not pressure. When the feedback is added over time, the energy is inserted into gas surrounding the star particle.  We also reran Run~1 between 19 and 27 Myr, but including star particles. In Run~1S, each star particle represents $\sim$50 M$_{\odot}$ of stars, and for Runs 4 and 5, each star particle represents $\sim$130 M$_{\odot}$ of stars.
 
In all the calculations there is uncertainty in the efficiency parameter, $\epsilon$, due to the IMF, any overcooling (though this is likely to be minimal with our resolution), and uncertainty in $M(H_2)$. Lastly in Runs~4 and 5, there is uncertainty because after the subsequent deposits of energy are always put into a fixed number of particles (30) compared to the first deposit of energy, where we simply insert energy within a smoothing length. 
\begin{table}
\centering
\begin{tabular}{c|cc|c|c|c|c|c}
 \hline 
Run & SFE & Stochastic? & Input over & Star & Time \\
& & & time? & particles &(Myr) \\
 \hline
1 & 0.4 &  N & N & N & 35\\
2 & 0.1 & N & N & N & 20\\
3 & 0.025 & Y & N & N & 20\\
4 & 0.15 & Y & Y (2Myr) & Y & 20\\
5 & 0.15 & Y & Y (5Myr) & Y & 20\\
1S & 0.4 &  N & N & Y & 8\\
\hline
\end{tabular}
\caption{Table of simulations shown in this paper. The middle columns provide information about the stellar feedback scheme, the star formation efficiency (SFE), whether the feedback is stochastic, and whether the feedback is input over time or instantaneous. In all the simulations, the star formation efficiency should not be considered exact (see text, particularly for Run~3). Run~1S is the same as Run~1, but includes star particles.}
\label{tab:runs}
\end{table}

Initially we aimed to run our simulations for several 10's Myrs, or 100 Myr, but the simulations already took substantial cpu time to reach a few 10's of Myrs. We note as well that other work has generally been less ambitious, Van Loo et al. 2013 only run their resimulations for $\sim$ 10 Myr. Here we run our simulation with $\epsilon=0.4$ for 35 Myr, and our other simulations for 20 Myr. As we show in the results, there is not very much difference with the different feedback schemes. In particular Runs 2, 3, 4 and 5 are extremely similar. To save space, we focus predominantly on Runs 1, 2 and 5. Runs 3 and 4 show very little difference to Run~2. For Sections~3-6, the results for Runs~3 and 4 lie within the range of results with the other feedback schemes. Run~1S is only shown in Section~7 on star formation rates. Runs 3-5 largely emerged as tests for differences in how the feedback is implemented rather than showing different results, although Run~5 probably has the most realistic feedback prescription.

\subsection{Cloud identification}
To identify clouds, we utilise two clump--finding algorithms, both described in Dobbs et al 2014, submitted. We use the method described in \citet{Dobbs2013}, where we searched for cells above a given column density, primarily to test cloud properties at different  resolutions. This method, which groups together all contiguous groups of cells above a certain column density, and defines these cells as a cloud, can readily be applied in exactly the same way at different resolutions, and the cell size can also be adjusted to compare different resolutions within a simulation. Results with this algorithm are shown in Section~3.2.1. 

In Dobbs et al, 2014 submitted., we show that this algorithm is not so robust when considered over different timeframes, and tends to produce quite blocky structures. Furthermore this is a 2D algorithm, and as we start to resolve all the vertical structure of the disc, the need to find clouds in 3D becomes more important. Hence we also apply an algorithm, shown in Dobbs et al. 2014, submitted., where we identify particles over a certain (volume) density, and group together all those within distance of each other into a cloud. This second algorithm is a `friends of friends' approach, and naturally produces clouds in 3D. We use this algorithm to consider properties of clouds more generally in these simulations, and compare the different models. For the analysis presented, we take a threshold density of 100 cm$^{-3}$ and a required distance between particles of 2.5 pc.

\section{Structure of the ISM}
\begin{figure*}
\centerline{\includegraphics[scale=0.34, bb=350 190 600 600]{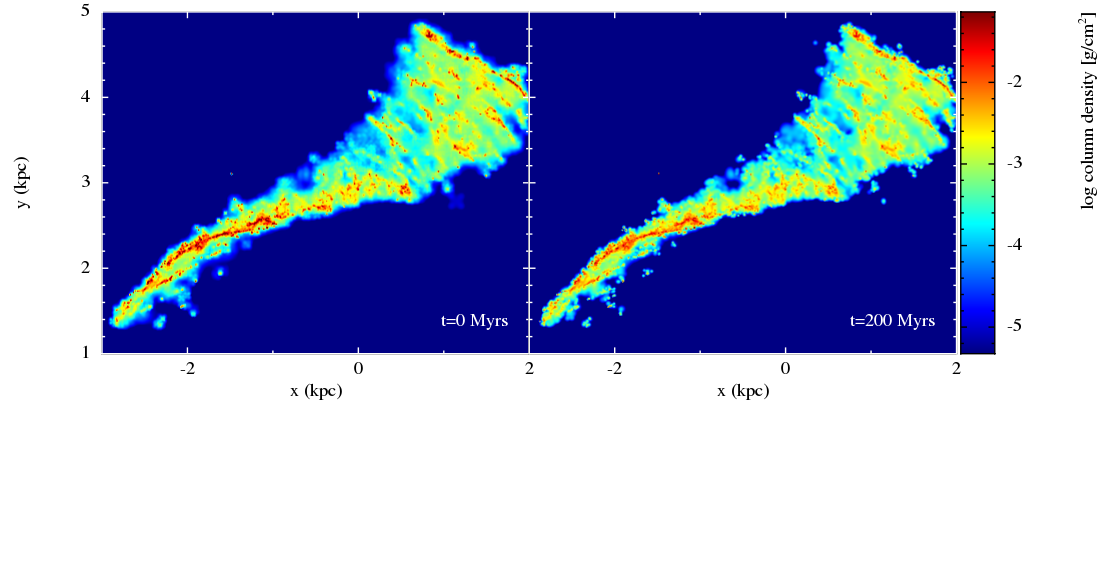}}
\centerline{\includegraphics[scale=0.34, bb=350 190 600 600]{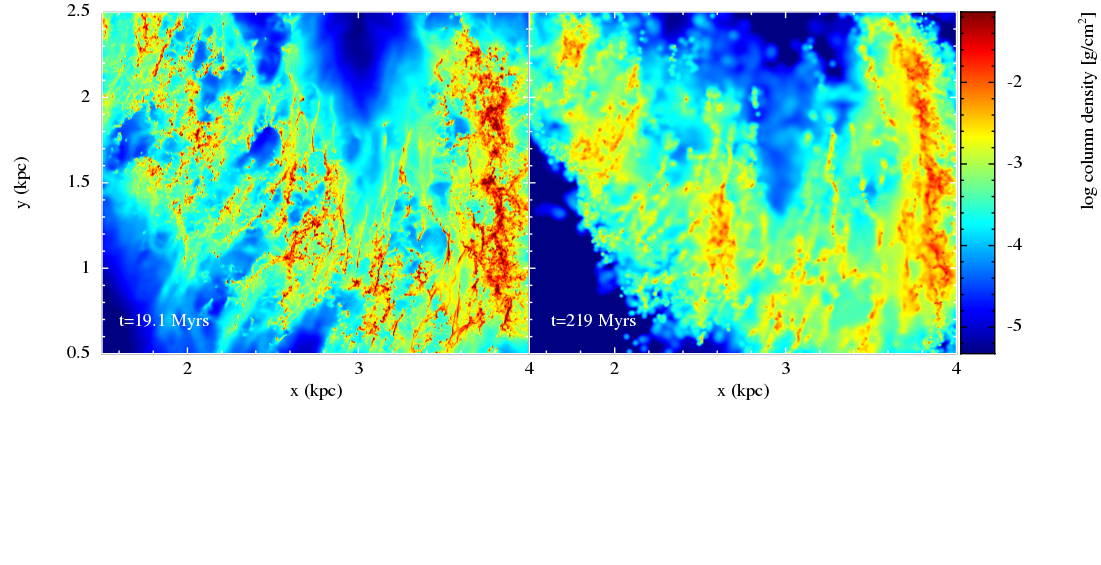}}
\centerline{\includegraphics[scale=0.34, bb=350 110 600 600]{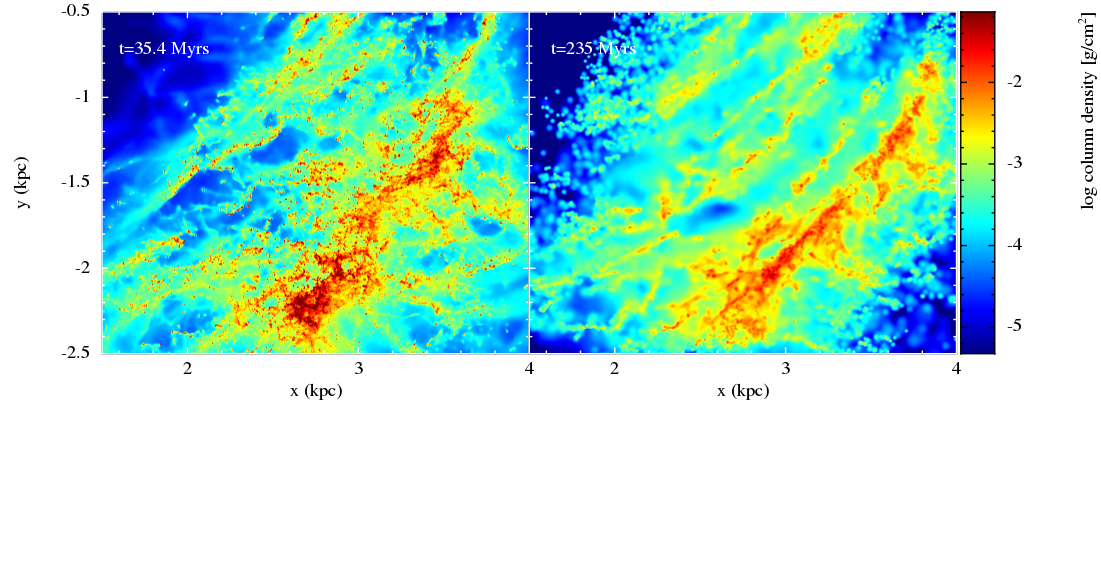}}
\centerline{\includegraphics[scale=0.3, bb=380 200 600 520]{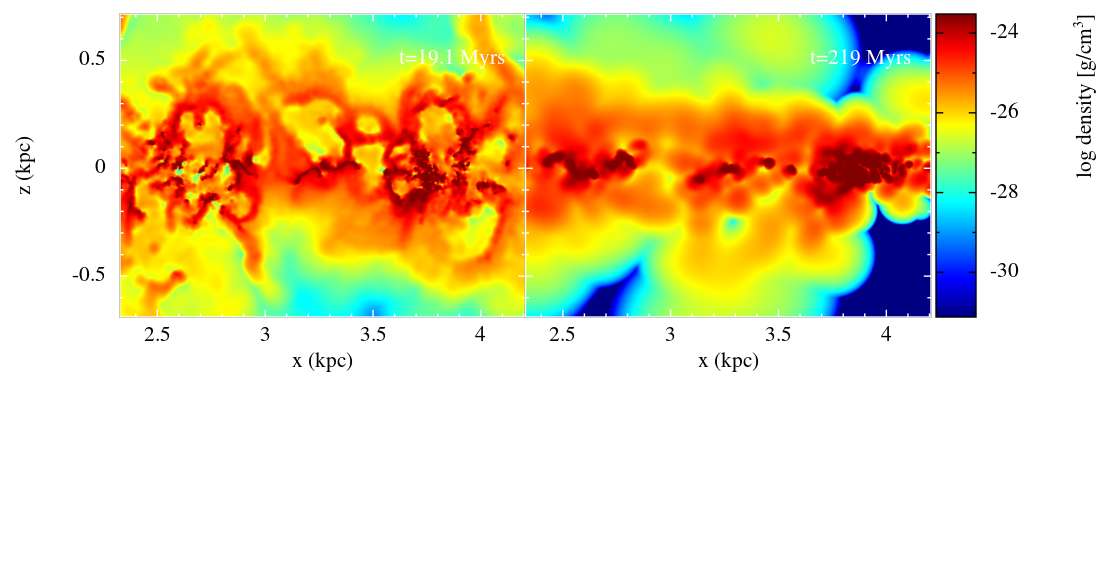}}
\caption{Comparison of structure between our high resolution resimulation (left) and the original global calculation (right). The maps for the global simulation are made using only those particles which were used to produce the initial conditions of the resimulation. The panels show the column density in the $xy$ (first, second and third panels) and a cross section in the $xz$ plane (lower panels) There is not a large difference between the global simulation and the resimulation at equivalent times - the main difference is simply that there is more structure in the resimulations, as would be expected. Shells in particular appear much better resolved.} Filamentary structure also appears to be finer in the resimulations, and more broken up.
\label{time_evolution}
\end{figure*}

\subsection{Structure of the ISM at high resolution vs low resolution}
In this section we discuss how the overall structure of the ISM compares in our global simulation and resimulations.
In Fig.~\ref{time_evolution} we compare the structure in our high resolution resimulation with $\epsilon=0.4$ (Run~1) with our original, whole galaxy calculation. As we discuss in the next section, the structure of the ISM is not that different with the different feedback prescriptions, so the comparisons made here are equally true for the other feedback prescriptions.  The panels in Fig.~\ref{time_evolution}  show 3 different times in the $xy$ plane and, one time in the $xz$ plane. Initially (top panels), there is minimal difference between the calculations, which is expected as the only change has been to increase the resolution. At later times, the large scale structure is still largely the same in both cases -- we can still pick out the same features in both simulations. However there is clearly more structure that we don't resolve in the whole galaxy simulation, that we do resolve in the resimulation. There is also a higher density contrast in the resimulation, with regions of higher column density, and also low column density. In particular clear bubbles, or shells due to feedback can be seen in the resimulation. We also see that at 19 Myr, there is considerable filamentary structure in the gas, the resimulation accentuating the structure already seen in the galaxy simulation. However by 35 Myr, when most of the gas is in the arm, filaments are less obvious in the gas. 

The lower panels of Fig.~\ref{time_evolution} show a cross-section in the vertical direction. Here there is a clearer difference between the simulations, with the higher resolution simulations showing many clear shells. Such features can be compared with shells seen in our and nearby galaxies (e.g. \citealt{Dawson2011,Dawson2013}).  Partly this particular simulation has a relatively high level of stellar feedback and produces more shells. But moreover the shells are better resolved. The global simulation appears to have a few low density holes in the gas, whereas in the resimulation there are much sharper shells, which are resolved by 100's of particles or more.

All our resimulations show clear filamentary structure. The filaments are much better resolved compared to the global simulations, which allows us to start analysing them (which will be considered further in future work). They have lengths of 100's pcs and widths of $\lesssim 10$ pc, considerably narrower than the global simulations so we note that the width may not have converged yet with resolution. As discussed in Smith et al. 2014, the narrow width of the filament is also likely to be due to the pressure in the ISM (both within the filament and outside), as well as shear. The filaments appear more numerous in the inter-arm regions, in agreement with observations by \citet{Ragan2014} of Galactic filaments of similar lengths. This could be due to the dynamics of the inter-arm regions compared to the spiral arms. In the inter-arm regions, the gas is subject to shear which helps create elongated structures. In the spiral arms, shear is minimal, whilst the gas is subject to compression predominantly in the direction perpendicular to the arm but also parallel to the spiral arm (in addition self gravity will lead to gas converging in the denser spiral arms). Alternatively the filaments could simply be more difficult to see in the spiral arms as they become compressed together. From Fig.~\ref{time_evolution} it is clear that the long filaments are the remnants of inter-arm spurs. Inter-arm spurs are in turn the remainders of massive spiral arm GMCs, which have been sheared out. In the spiral arms, these filaments are compressed, and again self gravity and increased stellar feedback from star formation likely makes this region more complex. A few filaments are found to be perpendicular to the plane, which are primarily associated with stellar feedback. Likewise vertical filaments found by \citet{Ragan2014} (GMF 20.0-17.9 and GMF 41.0-41.3) also appear to be associated with supernova remnants and regions of massive star formation.

\begin{figure}
\centerline{\includegraphics[scale=0.24, bb=450 160 600 600]{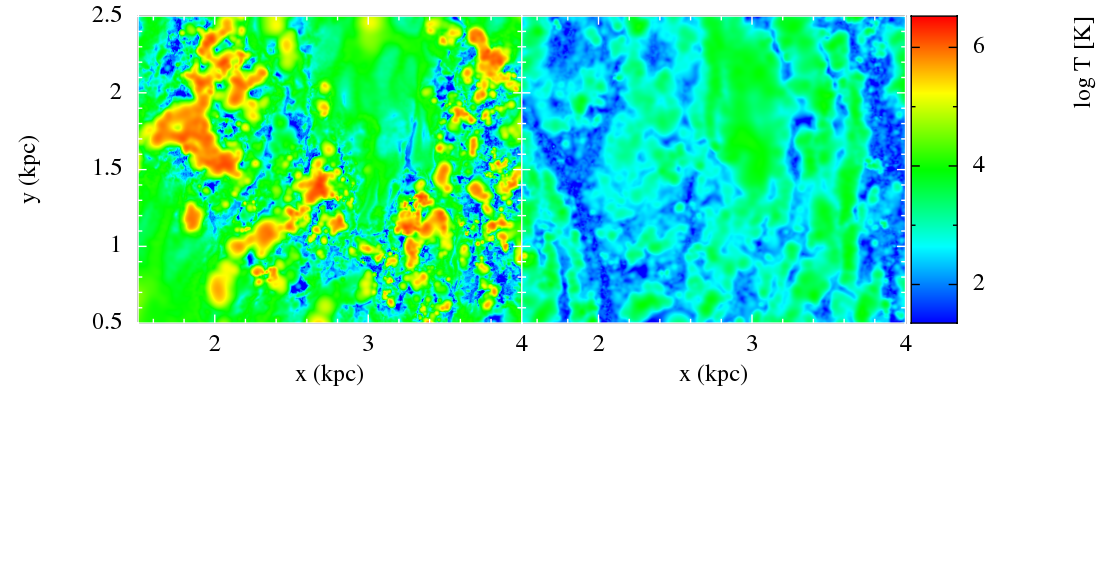}}
\centerline{\includegraphics[scale=0.24, bb=450 160 600 600]{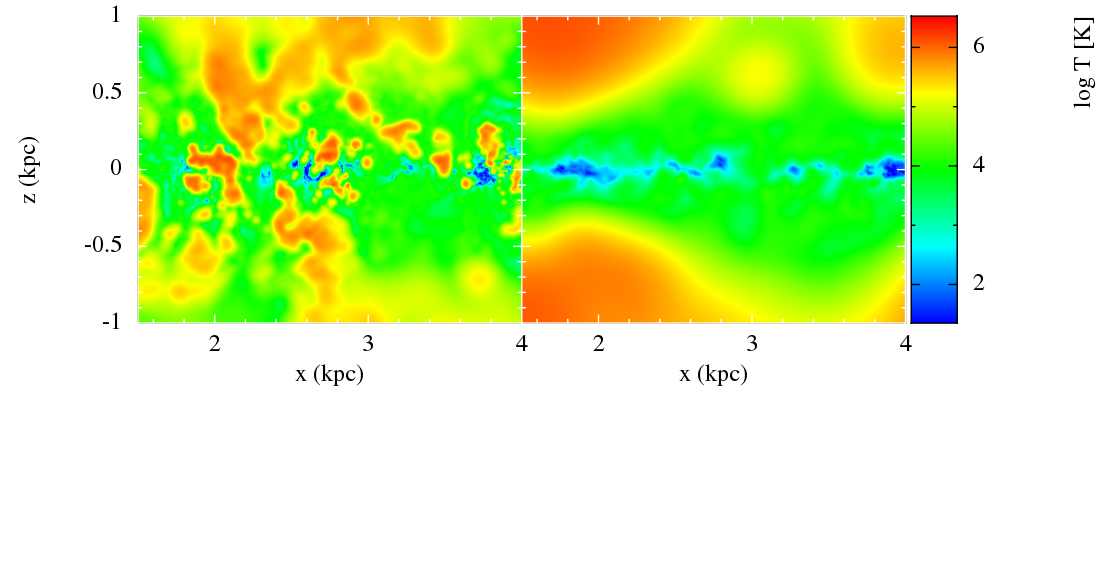}}
\caption{Comparison of the temperature in the resimulation (left) and original global calculation (right), at a time of 19 Myr in the resimulation. All panels show cross sections, the top panels show cross sections in the $xy$ plane, the lower panels show cross sections in the $xz$ plane. Hot gas is clearly more prominent in the resimulation compared to the global simulation, and much finer structure can again be seen in the resimulation.}\label{temperature}
\end{figure}

Fig.~\ref{temperature} shows cross sections of the temperature, in the $xy$ plane, and the $zx$ plane. Again the temperature shows much more structure in the resimulation compared to the whole galaxy simulation. There is much more hot gas in the resimulation, including in the mid-plane of the galaxy, whereas at low resolution, hot gas is confined to above and below the plane of the disc. In the resimulation, there are funnels of hot gas extending from the mid plane to above and below the plane of the disc. Again this suggests that we can resolve the local effects of feedback much better in the resimulations, and can better acquire significant volumes of relatively hot dense gas even in the mid plane of the galaxy, as expected from observational studies of the ionised ISM (e.g \citealt{Lockman1976,Dickey1990,Berk2006}).  

\subsection{Structure of the ISM with different feedback prescriptions}
\begin{figure*}
\centerline{\includegraphics[scale=0.3]{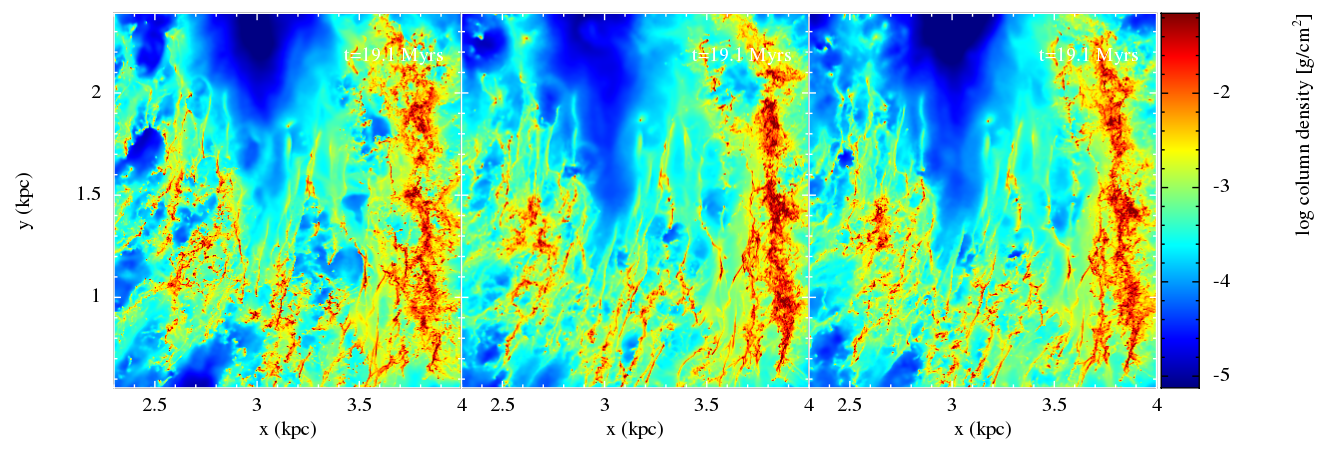}}
\centerline{\includegraphics[scale=0.28]{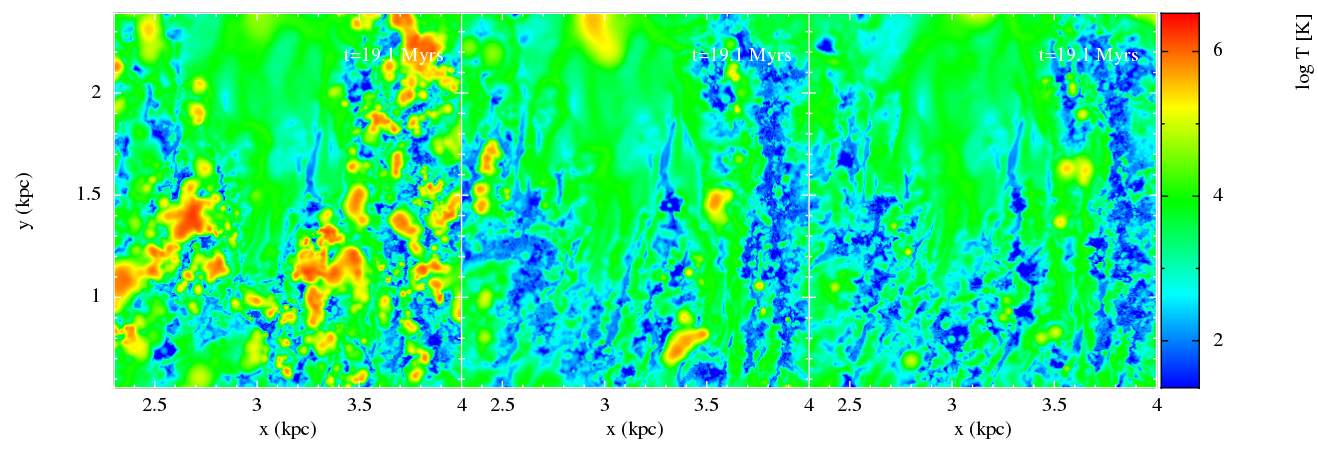}}
\centerline{\includegraphics[scale=0.3, bb=500 130 750 480]{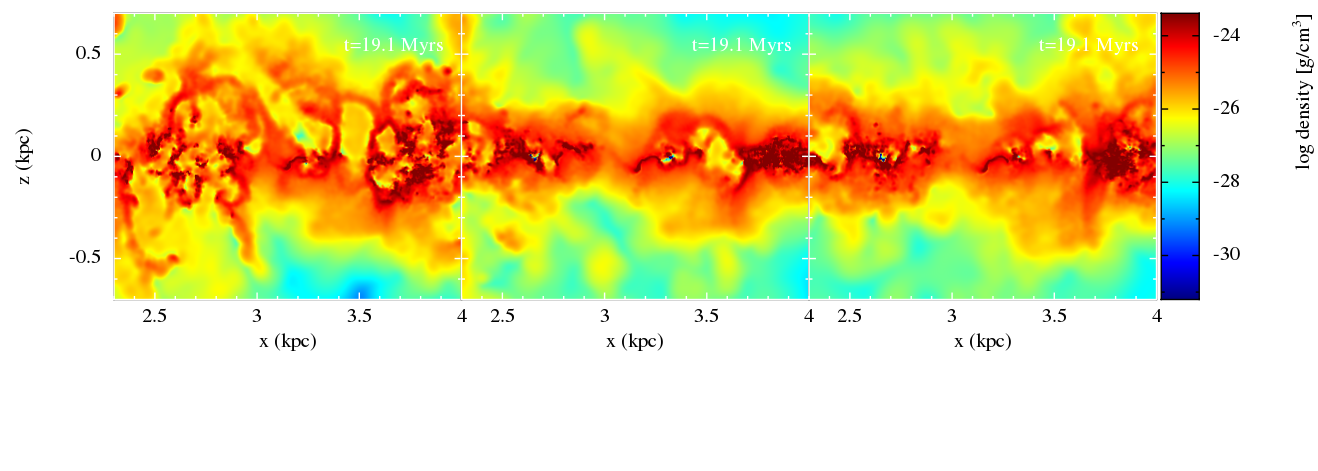}}
\caption{These panels compare different feedback prescriptions, the left panel shows the case where $\epsilon=0.4$ efficiency (Run 1), the centre $\epsilon=0.1$ (Run 2), and the right the case with feedback added over 5 Myr (Run 5). All the panels are shown at a time of 19 Myr. The structure in the plane of the disc (top panels) is very similar regardless of the feedback. The spiral arm appears slightly straighter with $\epsilon=0.1$, whilst there is more evidence of the impact of feedback with $\epsilon=0.4$ (e.g. at $\sim x=2.4$, $y=1.7$). The temperature (middle panels) shows a bigger difference. The temperature cross sections in the $xy$ plane (middle panel) indicate many more regions of hot gas in the $\epsilon=0.4$ model. The structure in the vertical plane (lower panels, again showing a cross section) is also quite different, with shells in the highest feedback case breaking out of the plane (left), but remaining more or less closed in the low feedback case (centre). The case with feedback added over time (right) maintains features of both other models, holes with hot gas, as well as dense gas confined to the mid plane, and a fairly highly structured spiral arm.}\label{fback_structure}
\end{figure*}

In this section we compare the structure of the ISM for the different feedback prescriptions implemented.
Fig.~\ref{fback_structure} shows column density plots for our feedback prescriptions with $\epsilon=0.1$ and 0.4 (Runs 1 and 2) and where feedback is added over 5 Myr (Run 5). Results from the stochastic prescription (Run 3) and Run~4 are similar to the case with $\epsilon=0.1$ (Run 2) and Run~5. The top panels show the structure in the $xy$ plane. There is surprisingly little difference between the varying levels of feedback.  The reason for this is probably that the feedback is most effective in the inter arm regions, where it shapes the bubbles in the ISM, but has less effect on the denser gas. The main difference is that the spiral arm has more structure in the case with higher feedback (Run 1), and that holes or bubbles in the gas are more obvious. The temperature maps (centre panels), which are cross-sections in the plane of the disc, show a larger difference. With $\epsilon=0.4$ compared to $\epsilon=0.1$ there is much more hot gas. With lower feedback, there is still hot gas, but it tends to reside above and below the plane of the disc (though note that there is still more hot gas on the mid-plane of the disc compared to the global low resolution simulation). 

The third panels in Fig.~\ref{fback_structure} show cross-sections of the density in the vertical direction.  With the higher feedback prescription (Run 1, left), the shells bear some resemblance to the models of \citet{Kim2011}, but are smoother. All models show clear shells and bubbles in the disc.  With the highest feedback prescription (Run 1), it is evident that the hot gas has often escaped out of bubbles, out of the plane of the disc. The models with feedback added over time (Runs 4 and 5) tend to retain structure around $z=0$, but are still able to push some gas to large $z$. Generally the models with feedback inserted over time manage to retain dense structures and the effects of feedback such as holes simultaneously better compared to the other models, where low or high feedback means one or the other tends to be prevalent, though this tends to be clearest in Run~4 and at earlier times. 

\begin{figure}
\centerline{\includegraphics[scale=0.4]{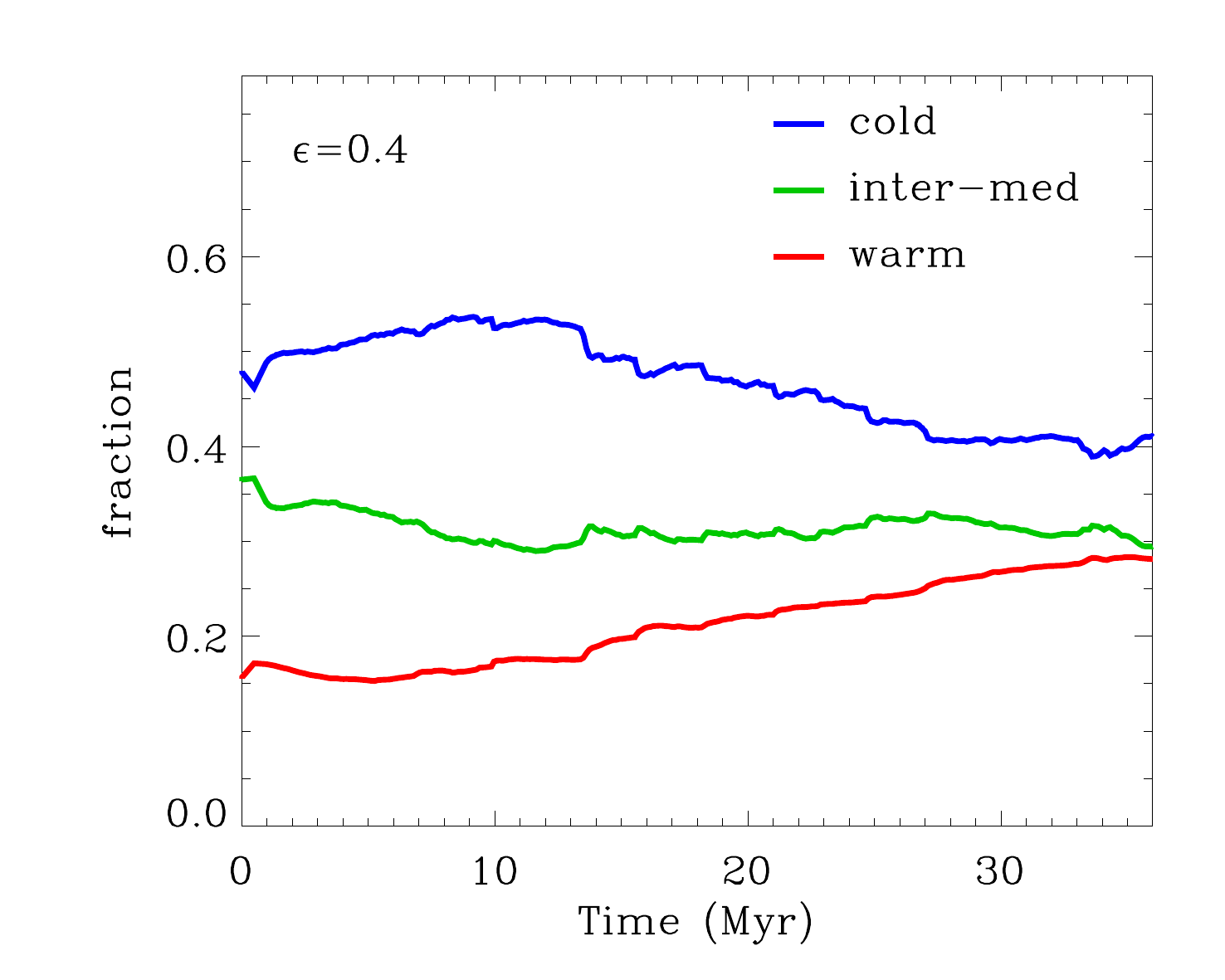}}
\centerline{\includegraphics[scale=0.42]{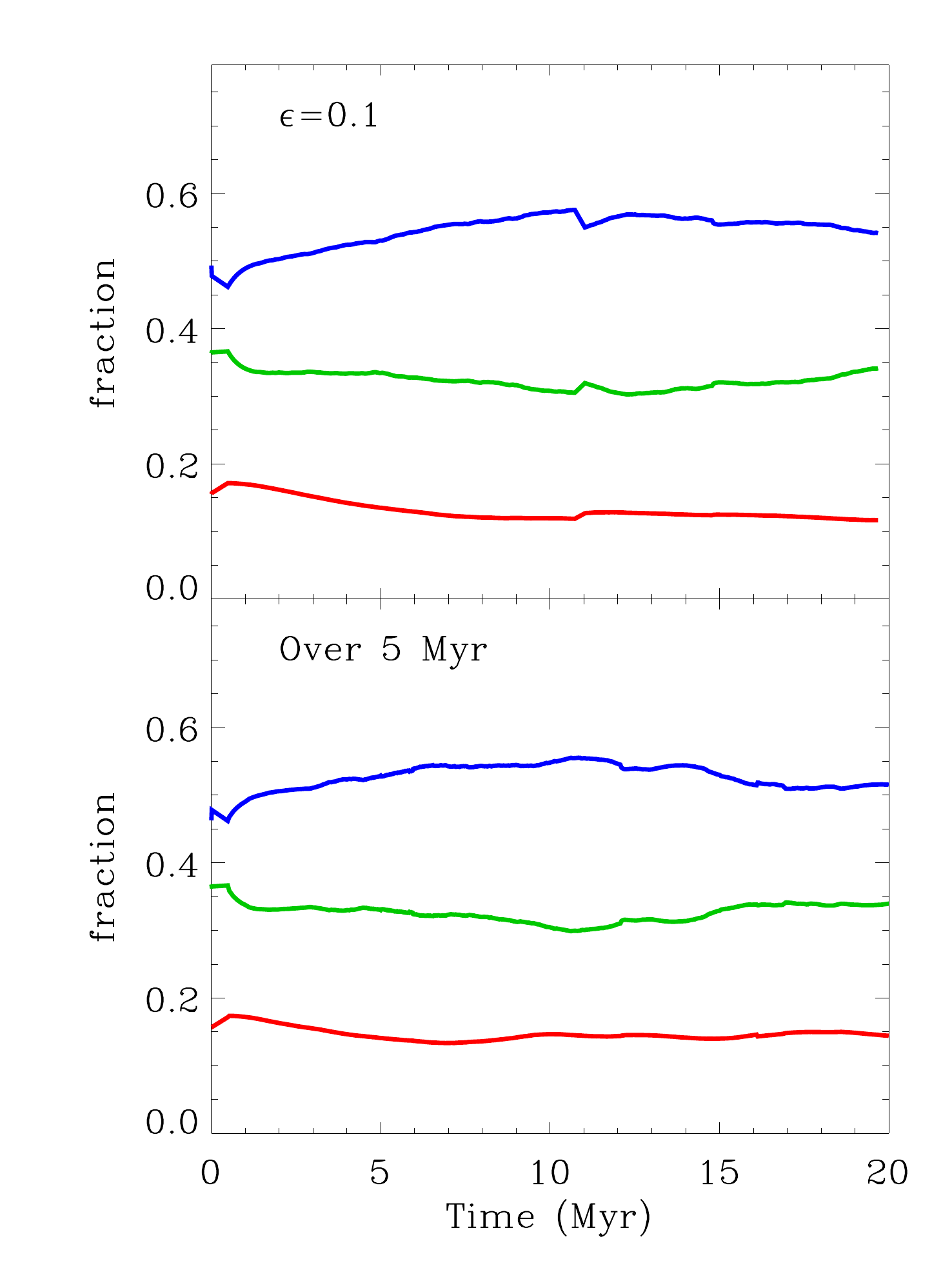}}
\caption{Comparison of the (mass-weighted) temperature evolution for the different feedback cases. With $\epsilon=0.4$ (top) the amount of cold gas decreases whilst the amount of warm gas increases. Conversely with $\epsilon=0.1$ (centre) the amount of cold gas increases, whilst the amount of warm gas decreases. With feedback added over time (lower panel) the distribution does not change much but the simulation contains more cold and warm gas simultaneously. For all cases there is not a significant departure from the initial temperature distributions over the course of the simulation. The jump in the centre panel represents a more intense period of feedback.}\label{temp_evol}
\end{figure}

As the main difference between the different feedback schemes appears to be the temperature, we show the temperature distribution explicitly in Fig.~\ref{temp_evol}. We divide gas into 4 regimes: cold ($<$ 150 K), intermediate (150 -- 5000 K), warm (5000 -- $5\times 10^5$ K) and hot  ($>5\times 10^5$ K, not plotted in Fig.~\ref{temp_evol}). As expected there is most cold gas in the case with lowest feedback ($\epsilon=0.1$, middle panel) and less warm gas.   The model with feedback inserted over time (Run~5) also contains around 50 \% cold gas (lower panel), and a slightly higher proportion of warm gas. This model shows very little change over time. Overall the temperature distribution does not change by more than 10 \% for all models. The amount of intermediate gas hardly changes. The amount of hot gas varies more in the models, reaching about 1\% for $\epsilon=0.4$, whereas this stays around $\lesssim 0.1$\% with the lower feedback. Fig.~\ref{temp_evol} seems to overall imply that temperature distribution is dominated by the overall galaxy conditions and structure, with some small changes over time according to the stellar feedback used in the resimulation. 

\section{Cloud properties}
\subsection{Comparison of different resolution simulations}
\begin{figure}
\centerline{\includegraphics[scale=0.28, bb=150 0 450 400]{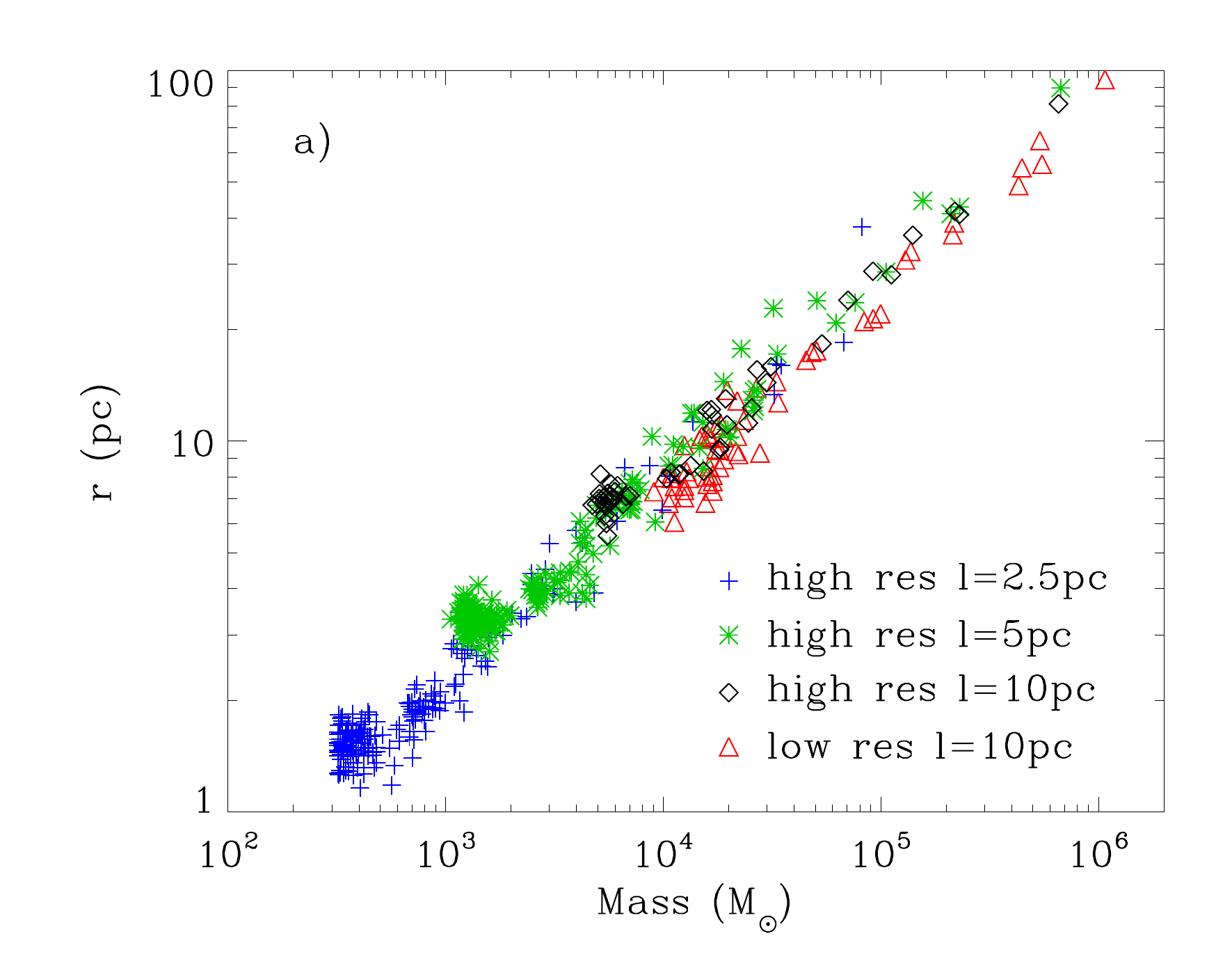}
\includegraphics[scale=0.28, bb=0 0 400 400]{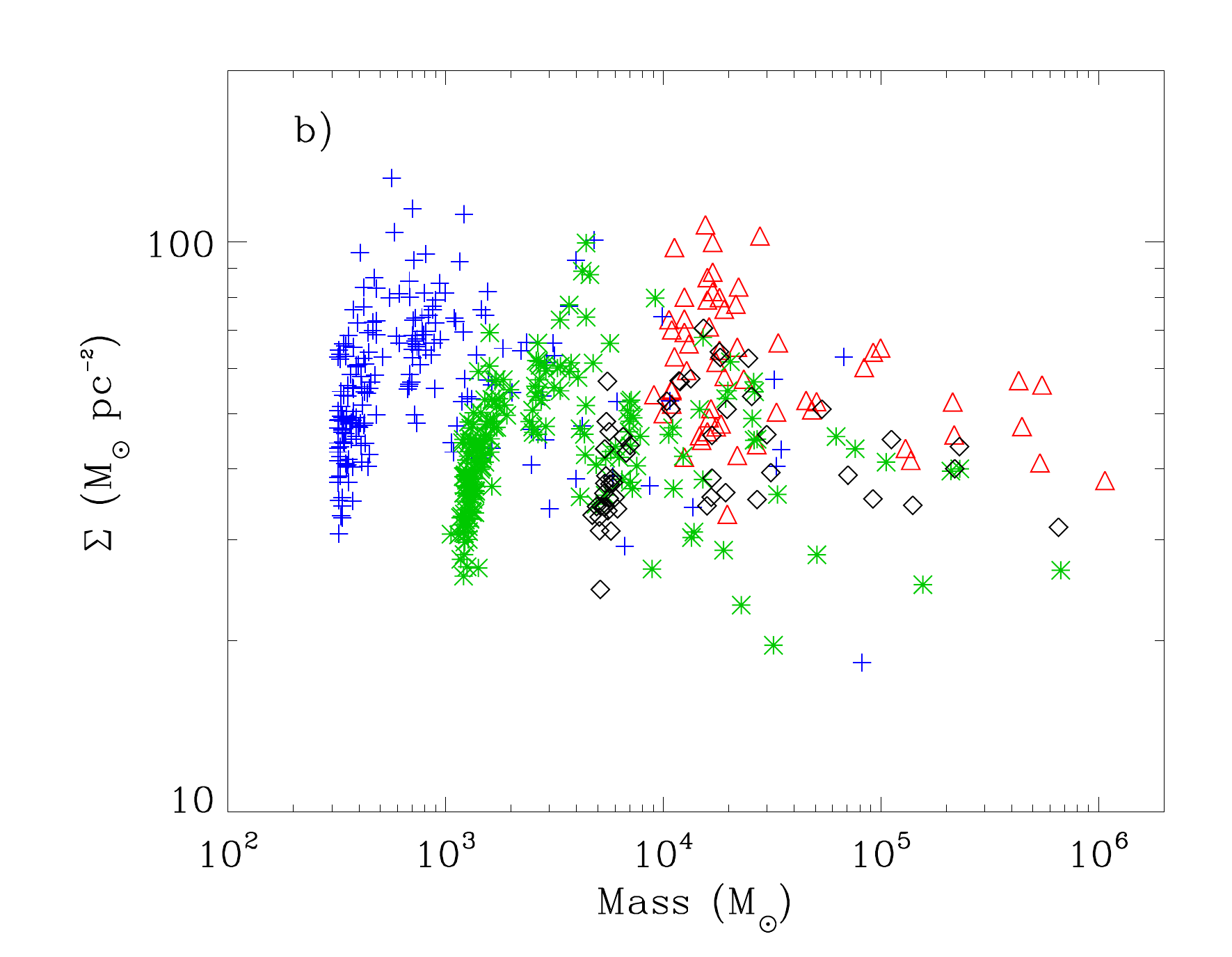}}
\centerline{\includegraphics[scale=0.28, bb=150 0 450 400]{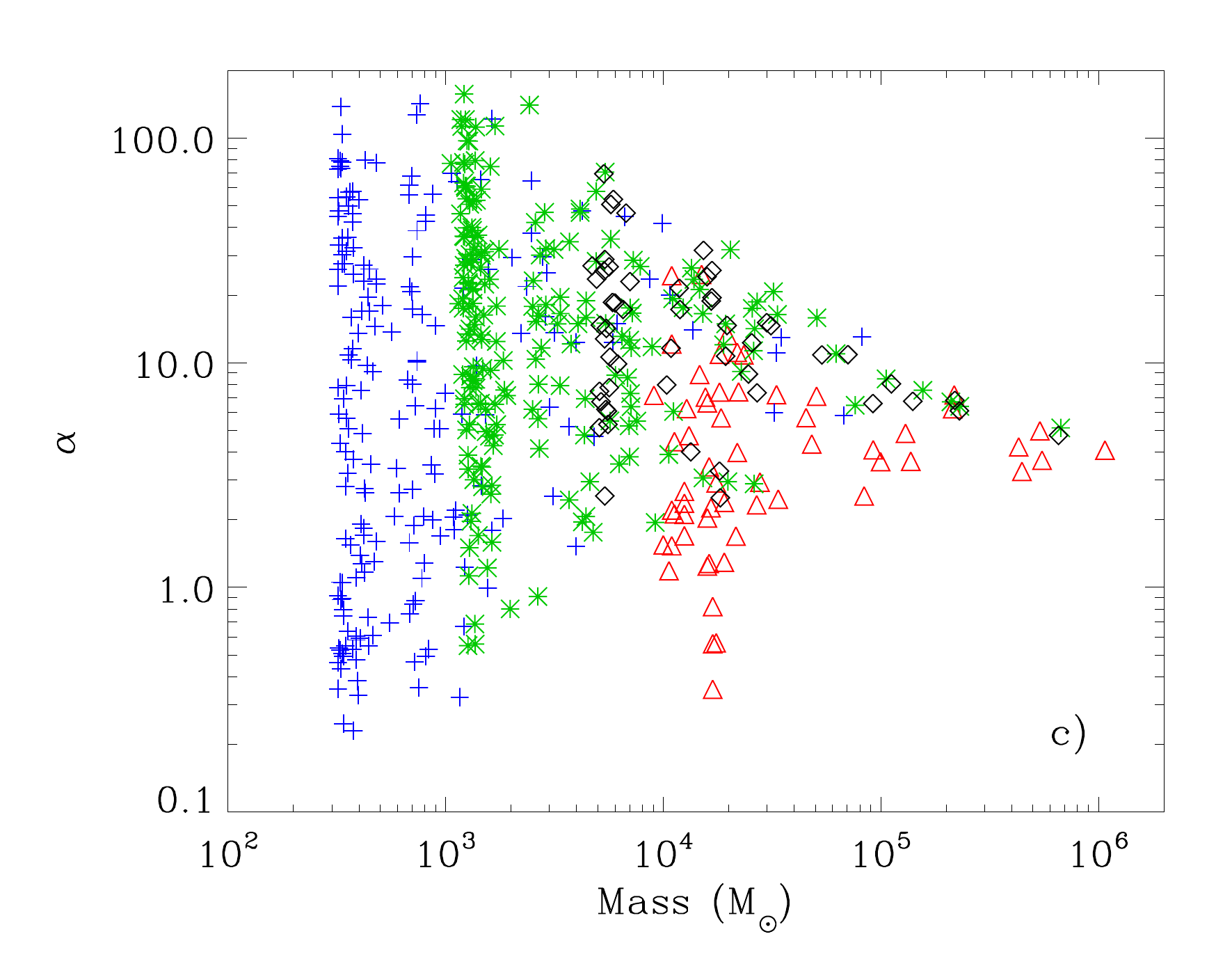}
\includegraphics[scale=0.28, bb=0 0 400 400]{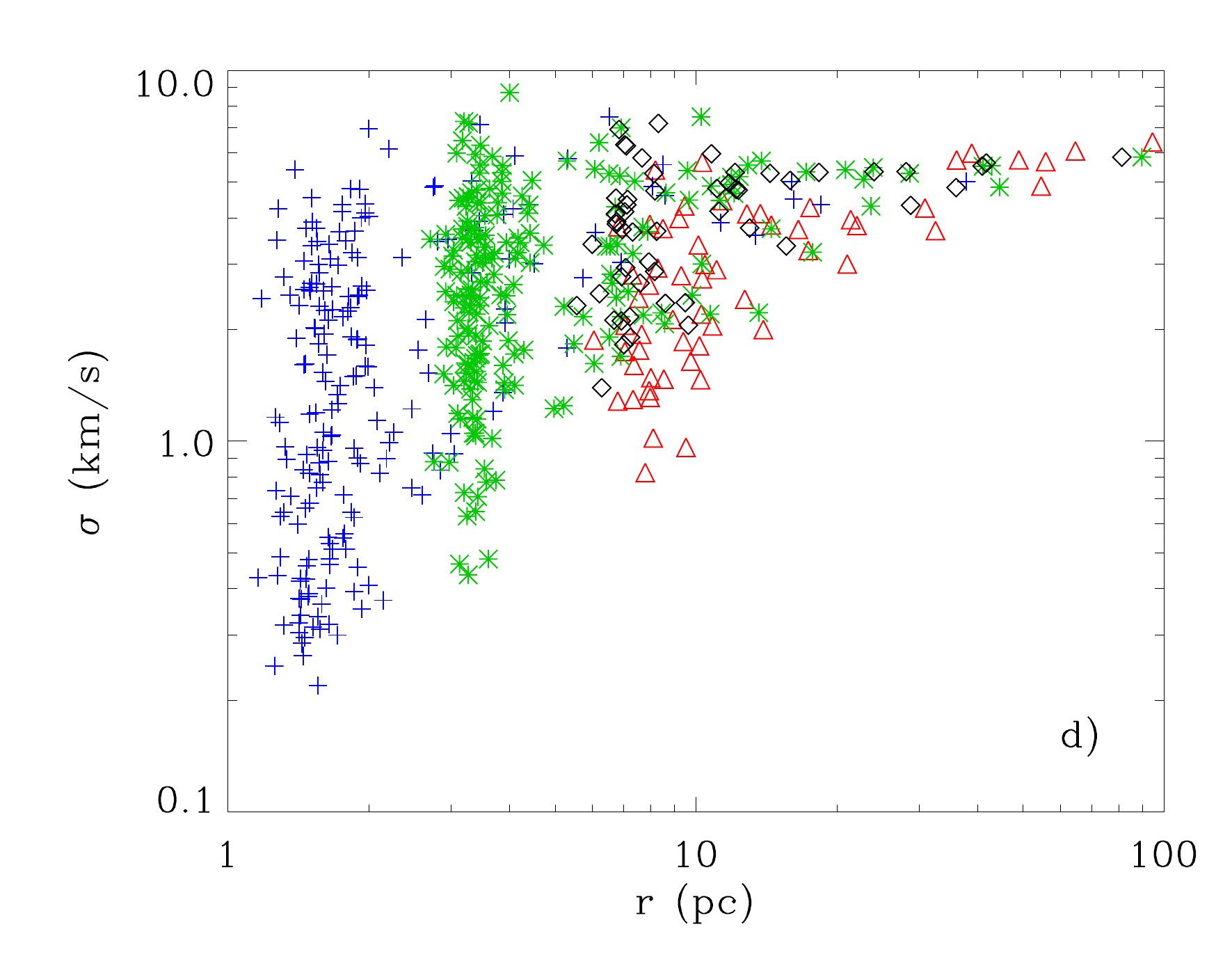}}
\caption{These panels, labelled a-d, show a comparison of cloud properties for the global simulation and the resimulation with $\epsilon=0.1$ (Run~2, red and black points) and for different resolution clump--finding algorithms (red, green and blue points). There is overall agreement at different resolutions, however some of the velocity dispersions appear to be low in poorer resolved clouds in the global, low resolution simulation.} 
\label{clouds_res}
\end{figure}

\begin{figure}
\centerline{\includegraphics[scale=0.38, bb=150 20 550 350]{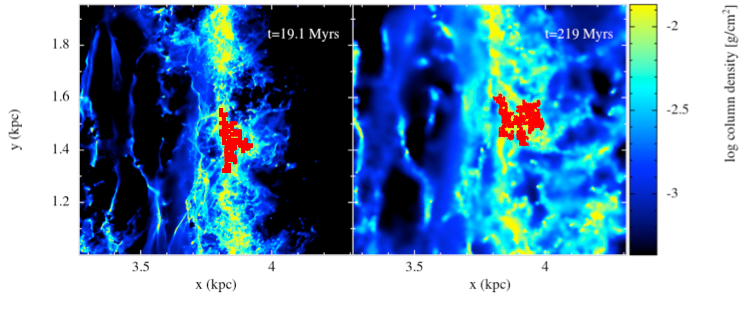}}
\caption{Equivalent clouds are shown for the global simulation (right) and the high resolution resimulation Run~2 (left), using the same clump finding algorithm.}\label{cloud_example}
\end{figure}

In this section we compare the properties of clouds in our original global calculation, and resimulations. We identify clouds at a time frame of 19 Myr in the high resolution simulation, and at the corresponding time of the global simulation (219 Myr), restricting the area of the global simulation to that roughly corresponding to the re-simulation. For this exercise we use the model with $\epsilon=0.1$ (Run 2). This is because this model most resembled the global simulation, in that the feedback prescription was instantaneous, and the feedback was less effective producing the best agreement visually with the global simulation. We compare cloud properties between the two simulations in Fig.~\ref{clouds_res}, and also show examples of clouds found in each simulation in Fig.~\ref{cloud_example}. For the cloud properties we compare, we apply exactly the same clump--finding algorithm to both simulations, using the grid--based algorithm, taking a threshold column density of 75 M$_{\odot}$ pc$^{-2}$ and a cell size of 10 pc. We also show clouds identified with smaller cell sizes (5 pc and 2.5 pc) for the high resolution simulation, which extends the range of clouds masses to 100's of M$_{\odot}$.

Overall there is good agreement between the clouds selected in the high and low (global) resolution simulations (red and black points). Fig.~\ref{cloud_example} shows an example equivalent cloud in the global and high resolution simulations, indicating that the same features can be identified.  However there are some differences between the different resolutions.

Fig.~\ref{clouds_res}b shows the surface densities of the clouds found in the simulations. The surface densities of clouds are similar, within a factor of 2, in each simulation. This is not particularly surprising given that the clouds we select in the simulations do not contain a large degree of significantly denser gas\footnote{Observed clouds may also show similar surface densities given a low fraction of high density material \citep{Lada2010,Kauffmann2010,Beaumont2012}.} given our limit for imposing feedback. There is a small difference between the clouds in the resimulation versus the global simulation, those in the resimulation having slightly lower surface densities. This could be because the structure is finer in the resimulations so the mass and radius estimates are lower, but as the simulations are not completely the same it is difficult to be certain.

Fig.~\ref{clouds_res}c shows the virial parameters. The virial parameters tend to be slightly higher in the resimulation, probably because the stellar feedback is being added down to smaller scales. The main difference is that some of the less well resolved clouds in the global simulation (of $10^4-2\times 10^4$ M$_{\odot}$) clearly lie at lower $\alpha$ than any clouds in the high resolution simulation. Only this region of unreliable clouds (all of which are less than 100 particles) includes clouds with $\alpha<1$. A comparison of the other properties ($r, M$ and $\sigma$, see Fig.~\ref{clouds_res}) indicate that it is likely $\sigma$, the velocity dispersion, where the discrepancy lies, whereas the differences in radius and mass are only small (e.g. Fig.~\ref{clouds_res}a). This could reflect that there are not enough particles to correctly determine the velocity dispersion, and sources of energy below the resolution of the simulation are absent. 

Fig.~\ref{clouds_res} also shows cloud properties for the high resolution resimulation but with different grid cell sizes for the clump--finding algorithm (effectively the resolution of the clump--finding scheme). The cloud properties appear to be similar and independent of the cell size, except obviously lower mass clouds can be detected. 

\begin{figure}
\centerline{\includegraphics[scale=0.36]{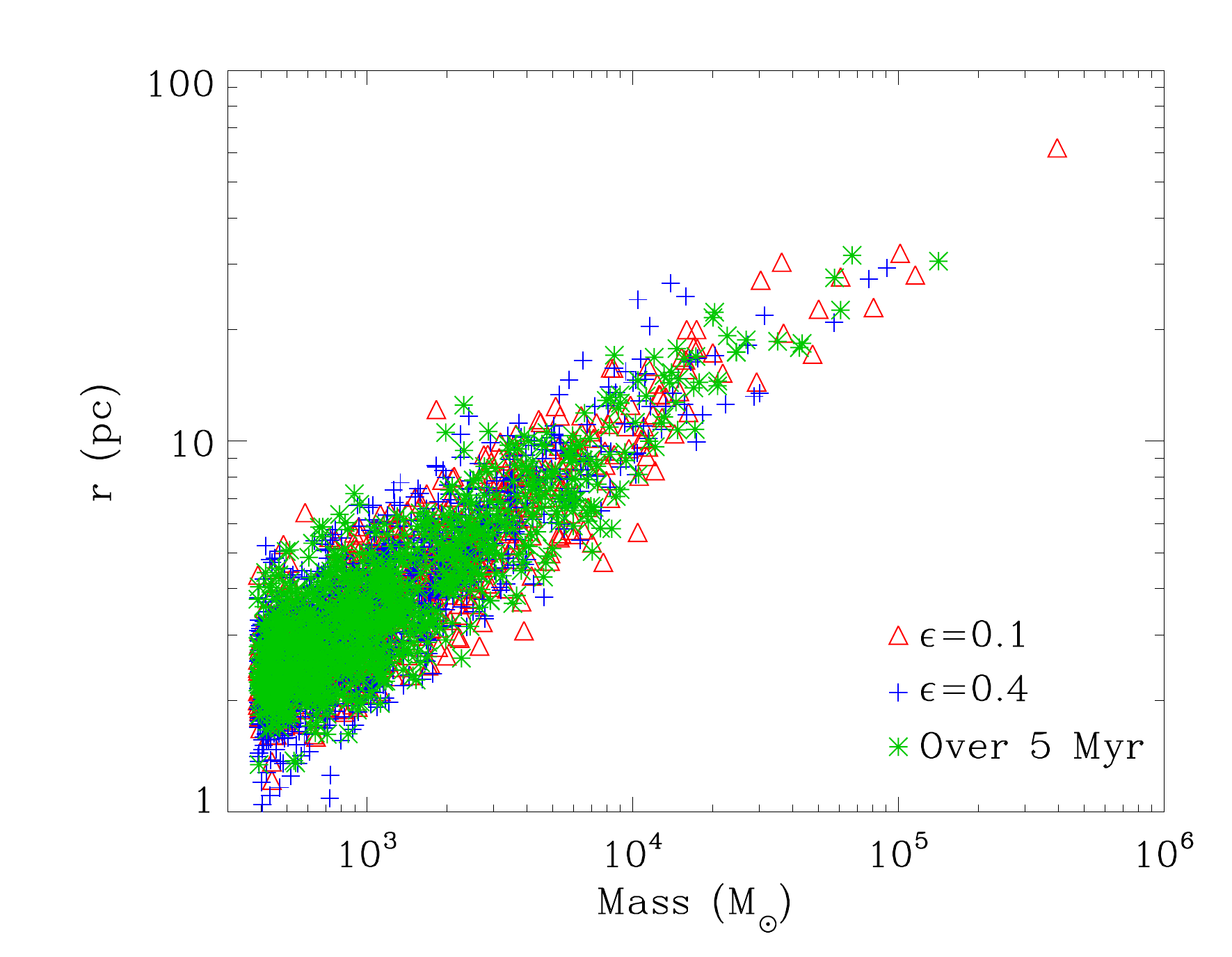}}
\centerline{\includegraphics[scale=0.36]{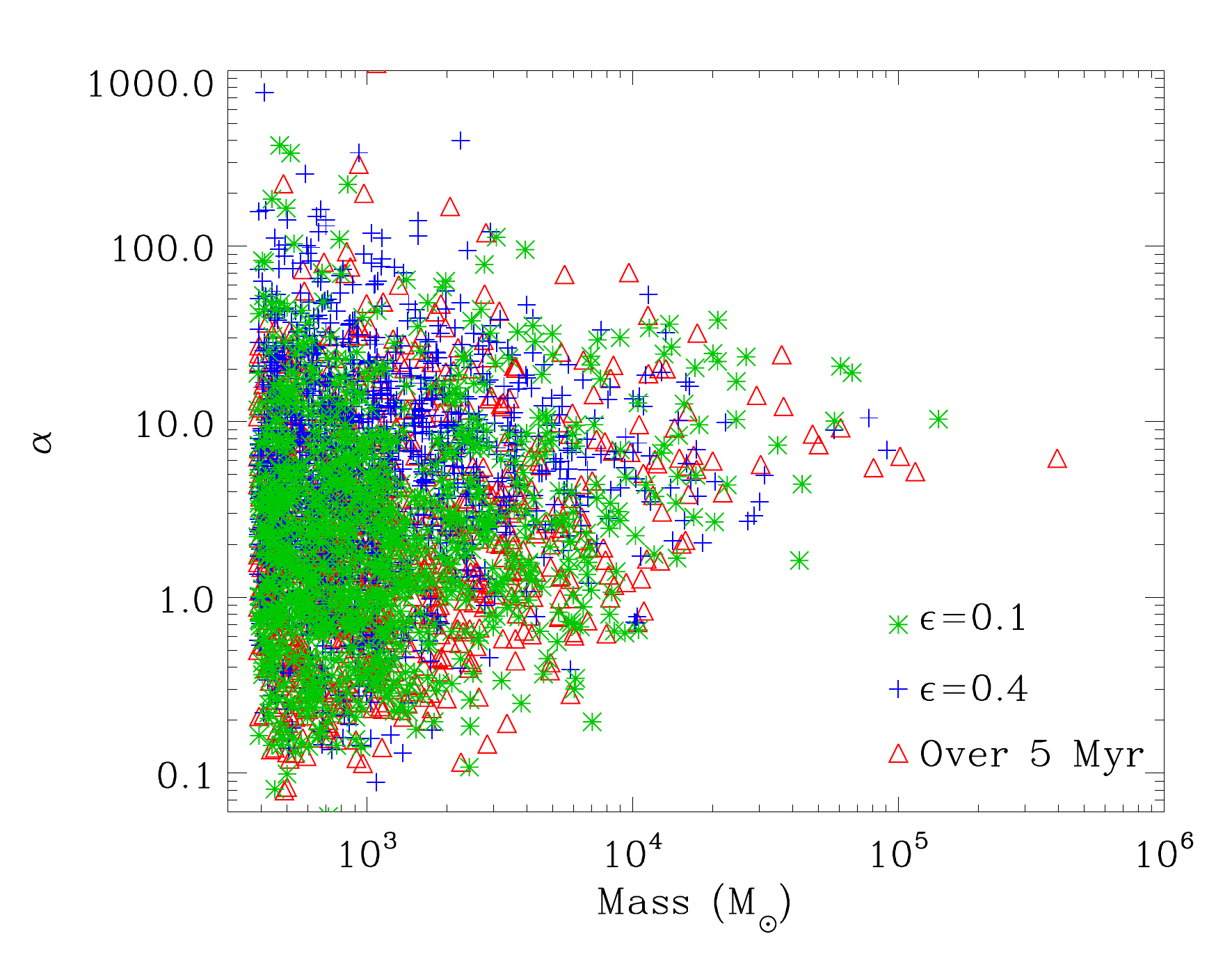}}
\centerline{\includegraphics[scale=0.36]{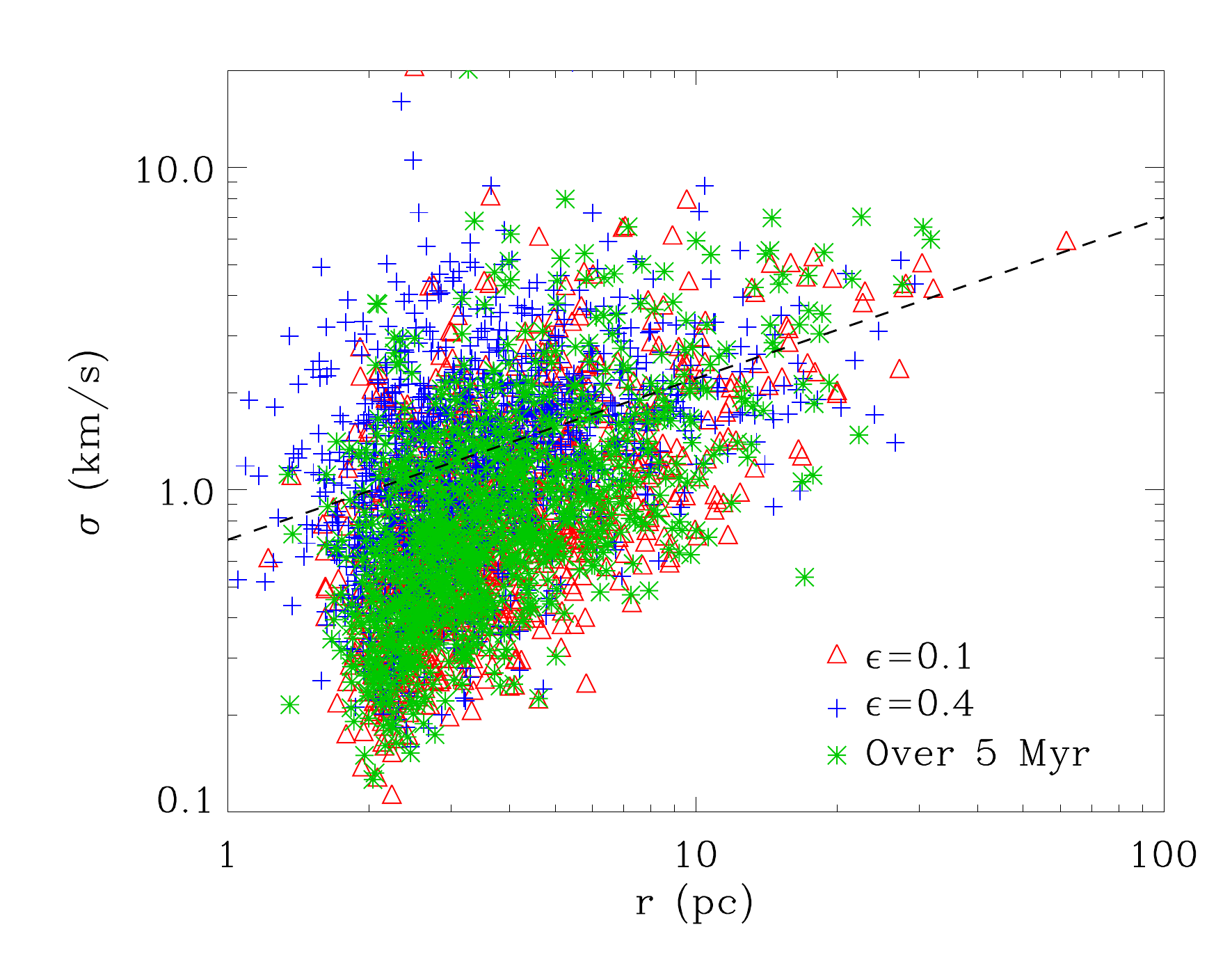}}
\caption{Cloud properties are compared for different feedback prescriptions (all for the high resolution resimulation). The panels show mass and radius (top), virial parameter (middle) and velocity dispersion (lower panel). The dashed line on the lower panel shows $\sigma=0.7 (r/1$pc$)^{0.5}$ km s$^{-1}$.}\label{clouds_fback}
\end{figure}

\subsection{Comparison of different feedback schemes}
In this section we use the second, 3D, `friends of friends' clump--finding algorithm. In all the following figures, only clouds with at least 100 particles are shown. This still allows us to explore cloud masses as low as 400 M$_{\odot}$. The results presented are similar when using either algorithm, as can be seen when comparing Figures~\ref{clouds_res} and \ref{clouds_fback}. The main difference is that there is slightly more scatter with the friends of friends approach, which tends to reflect the ability of this algorithm to pick out the densest gas (e.g. a gravitationally collapsing cloud or a long filament), without selecting background low density material.

We show various cloud properties in Fig.~\ref{clouds_fback}. Again for clarity we don't show the stochastic model (Run 3) or Run~4, but note that the properties of the clouds in these simulations are similar. Generally the properties appear to be similar regardless of the feedback prescription, and differences in the populations of clouds are hard to pick out. The mass--size relation (top panel) for the clouds is indistinguishable. The virial parameters (middle panel) are also similar, although the prescription with low feedback ($\epsilon=0.1$, Run 2) has notably more clouds at lower $\alpha$ compared to the other models. Clouds at low masses appear to exhibit  a large range of virial parameters, narrowing with increasing cloud mass. We find that the more massive clouds are surprisingly unbound, with virial parameters typically around 5--10. However this is partly a consequence of our initial conditions and section of the galaxy. For clouds over the original galaxy simulation \citep{Dobbs2013} there are some massive bound clouds, but there are relatively few across the whole galaxy, and we simply don't include many in our chosen section. We might expect more massive clouds $>10^6$ M$_{\odot}$ to be more bound (based again on \citealt{Dobbs2013}) but likewise there are no such clouds in this section. Run~4 (not shown) included a few more marginally bound clouds in the range $10^4-10^5$ M$_{\odot}$, probably the consequence of this model containing slightly more dense gas as well as effective feedback. In the third panel we show the velocity dispersion versus size, and there is some indication of a trend for all the models but there is quite a lot of scatter. The stochastic model (Run 3, not shown) showed the least scatter.

\begin{figure}
\centerline{\includegraphics[scale=0.35]{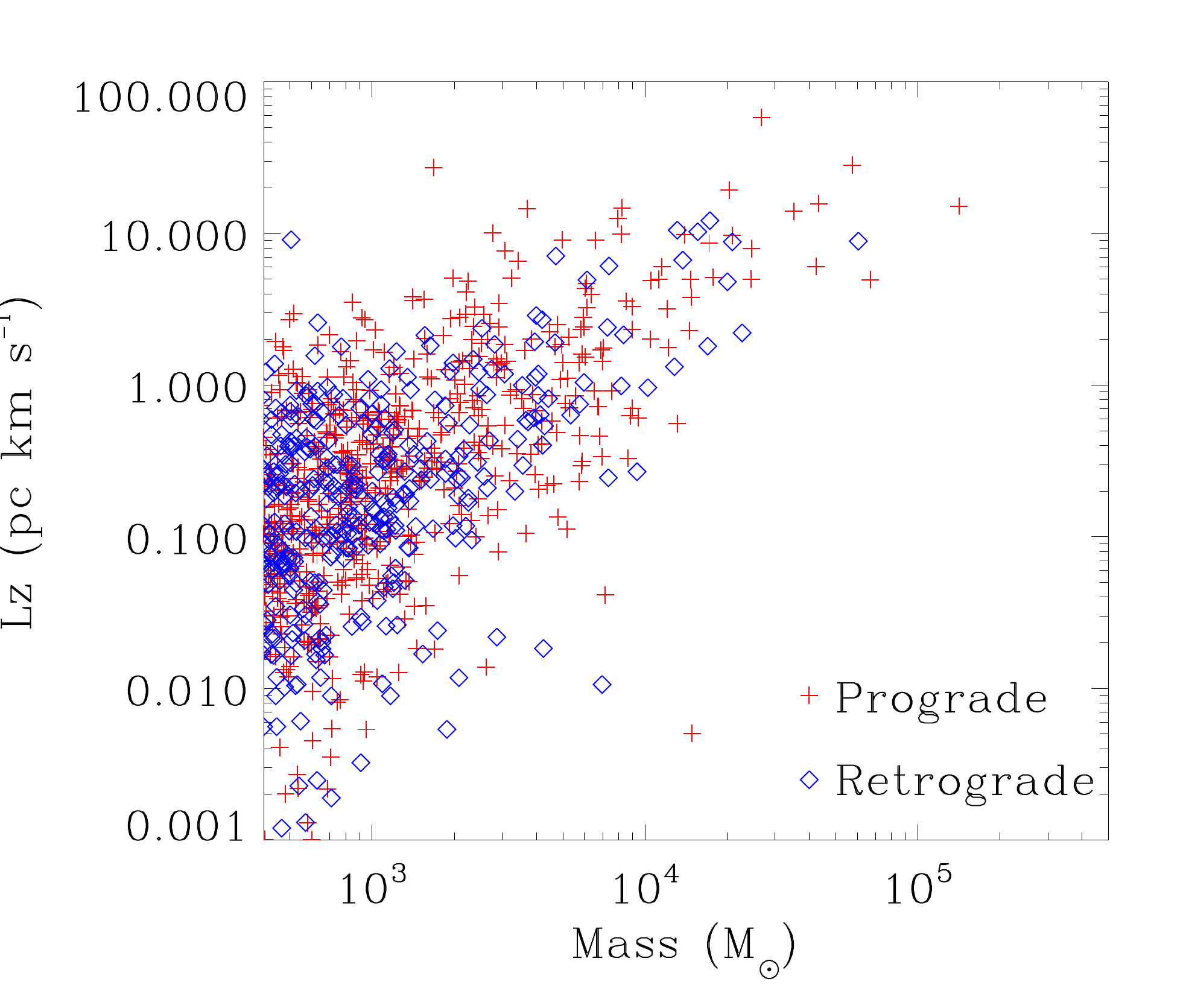}}
\centerline{\includegraphics[scale=0.35]{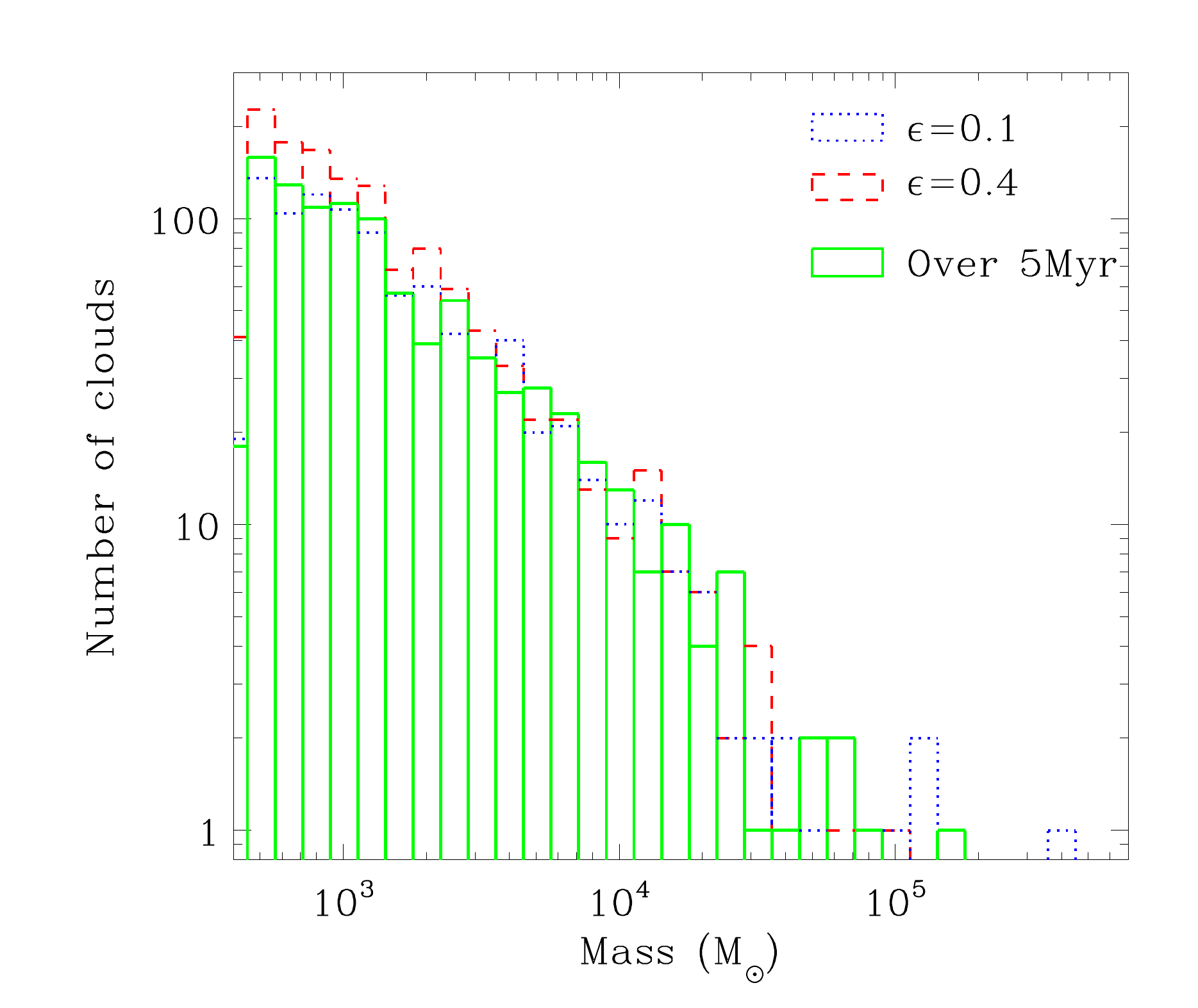}}
\caption{The angular momenta are shown for clouds from Run~5 (top panel). Other models or criteria show similar results. The fraction of retrograde clouds is 47 \% in this case. The distribution of angular momenta is similar to \citet{Dobbs2013}. The mass spectra for the clouds in Runs 1,2 and 5 (with high and low feedback, and feedback spread over time) are shown in the lower panel.}\label{mom_spec}
\end{figure}

We also looked at cloud rotations, but as there were negligible differences between the different models, we only show the results from one simulation, with the prescription with feedback added over 5 Myr (Run 5) in Fig.~\ref{mom_spec} (top panel). The distribution agrees well with those of the global simulation \citep{Dobbs2013}, with 47 \% retrograde clouds here compared to 40\% in the global calculation. There are few clouds with angular momenta $<0.1$ pc km s$^{-1}$ and mass $>10^4$ M$_{\odot}$ in the high resolution resimulations but again this could reflect the far smaller number of clouds in this regime compared to the global simulation. Resolution may be significant as well though, since we are now resolving dynamics on much smaller scales which may contribute to the the clouds' angular momenta. The fraction of retrograde clouds was very consistent between the different feedback models, lying in the range 40-50 \%, and again is very consistent with previous global studies \citep{Dobbs2011new,Dobbs2013} and observations \citep{Rosolowsky2003,Imara2011a}. Similarly the fraction of retrograde clouds does not vary according to the clump--finding scheme adopted.

Finally in the lower panel of Fig.~\ref{mom_spec} we show mass spectra for the clouds, this time for Runs 1, 2 and 5. Again, there is little difference between the different feedback schemes, with the power spectra exhibiting a similar slope $dN/dM\propto \sim M^{\sim 2}$ in all cases. There is also no turnover evident at lower masses. The main difference is the maximum mass of clouds. With the lowest level of feedback ($\epsilon=0.1$, Run 1), there are a few more massive clouds, which would otherwise be broken up with the other feedback schemes. The mass spectrum for the run with feedback added over 5 Myr (Run~5) is again very similar to the models with instantaneous feedback.
 
\subsection{Internal properties of highly resolved clouds}
\begin{figure}
\centerline{\includegraphics[scale=0.32]{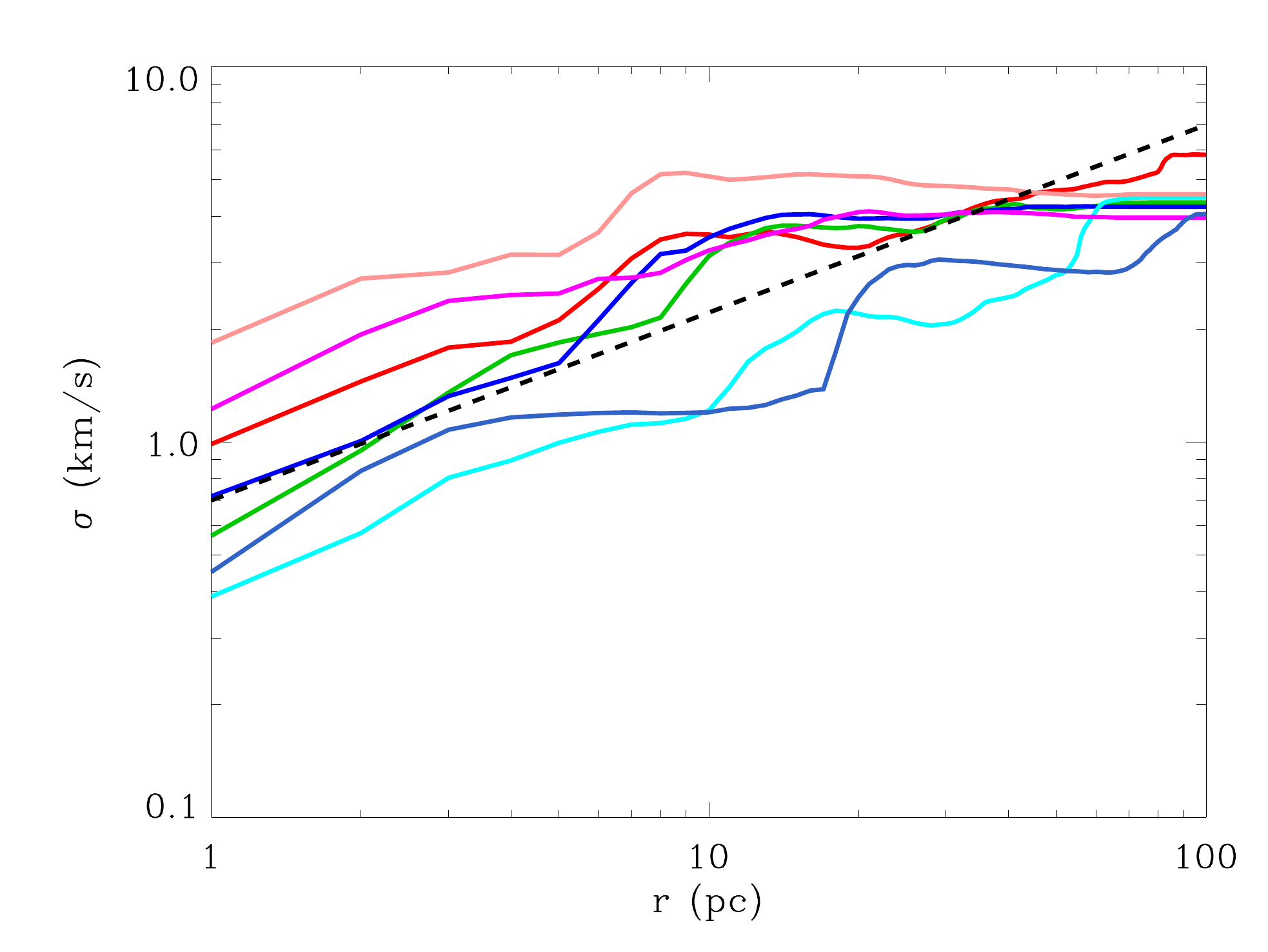}}
\centerline{\includegraphics[scale=0.32]{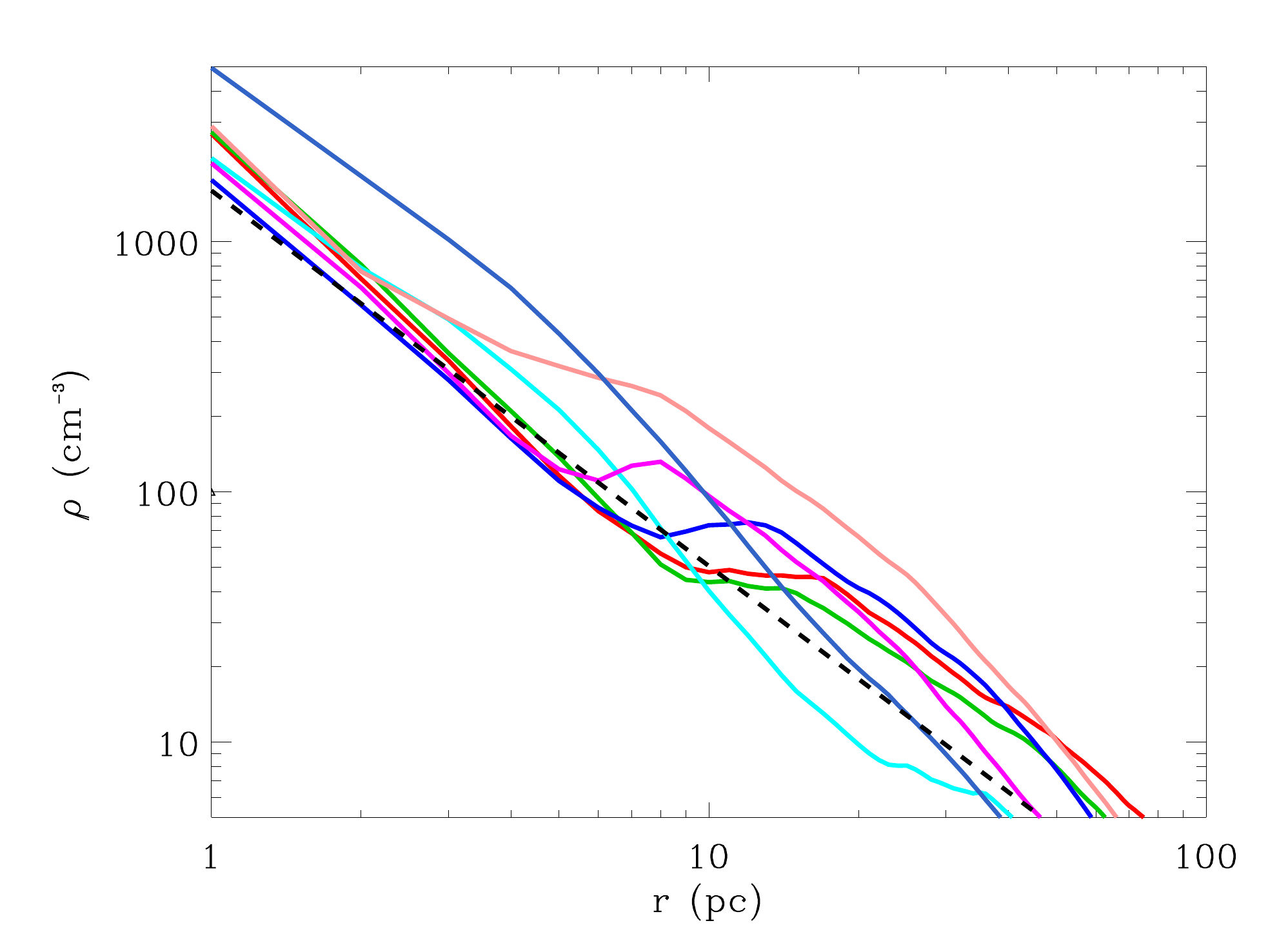}}
\centerline{\includegraphics[scale=0.3, bb=150 0 400 520]{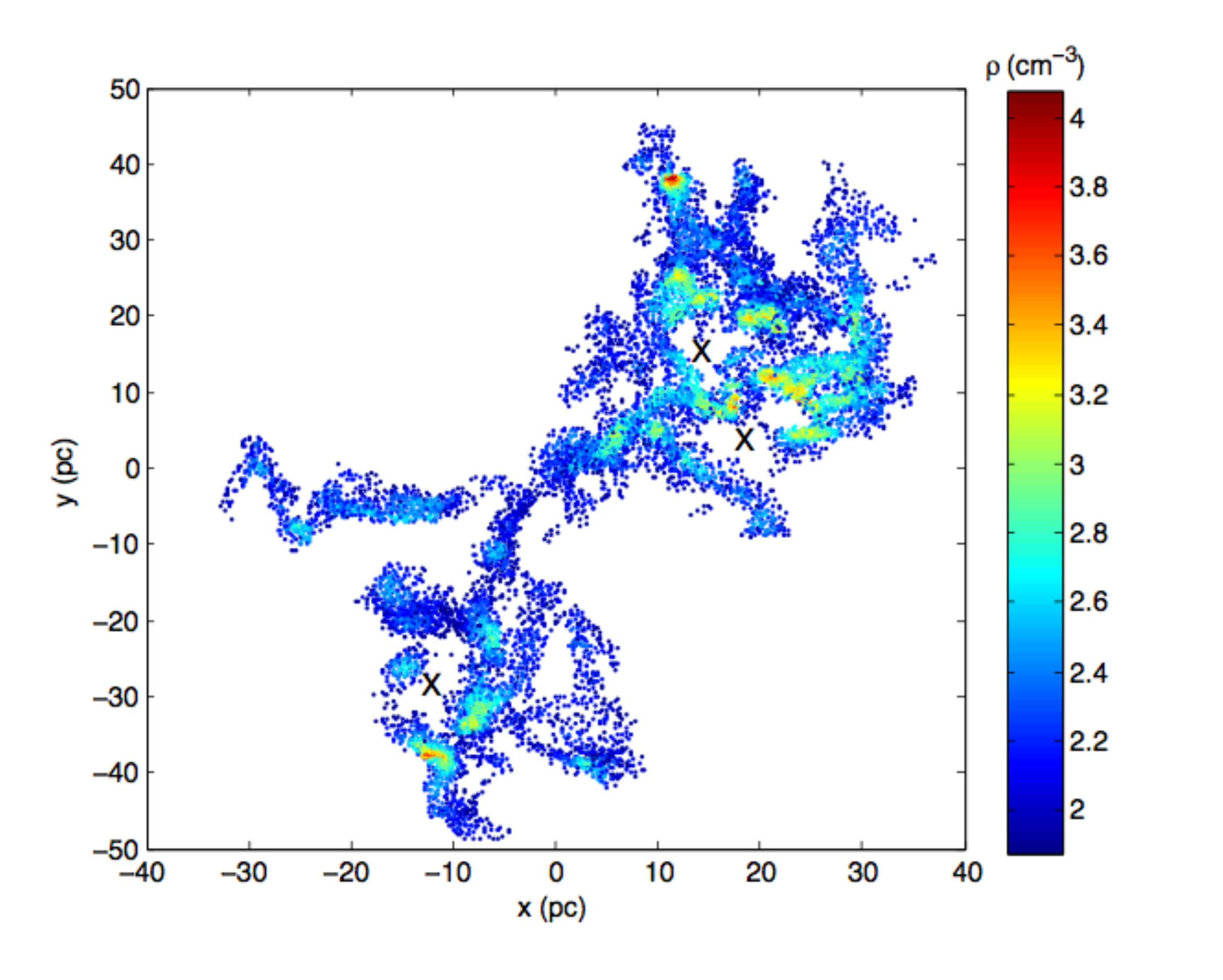}}
\centerline{\includegraphics[scale=0.3, bb=130 30 400 490]{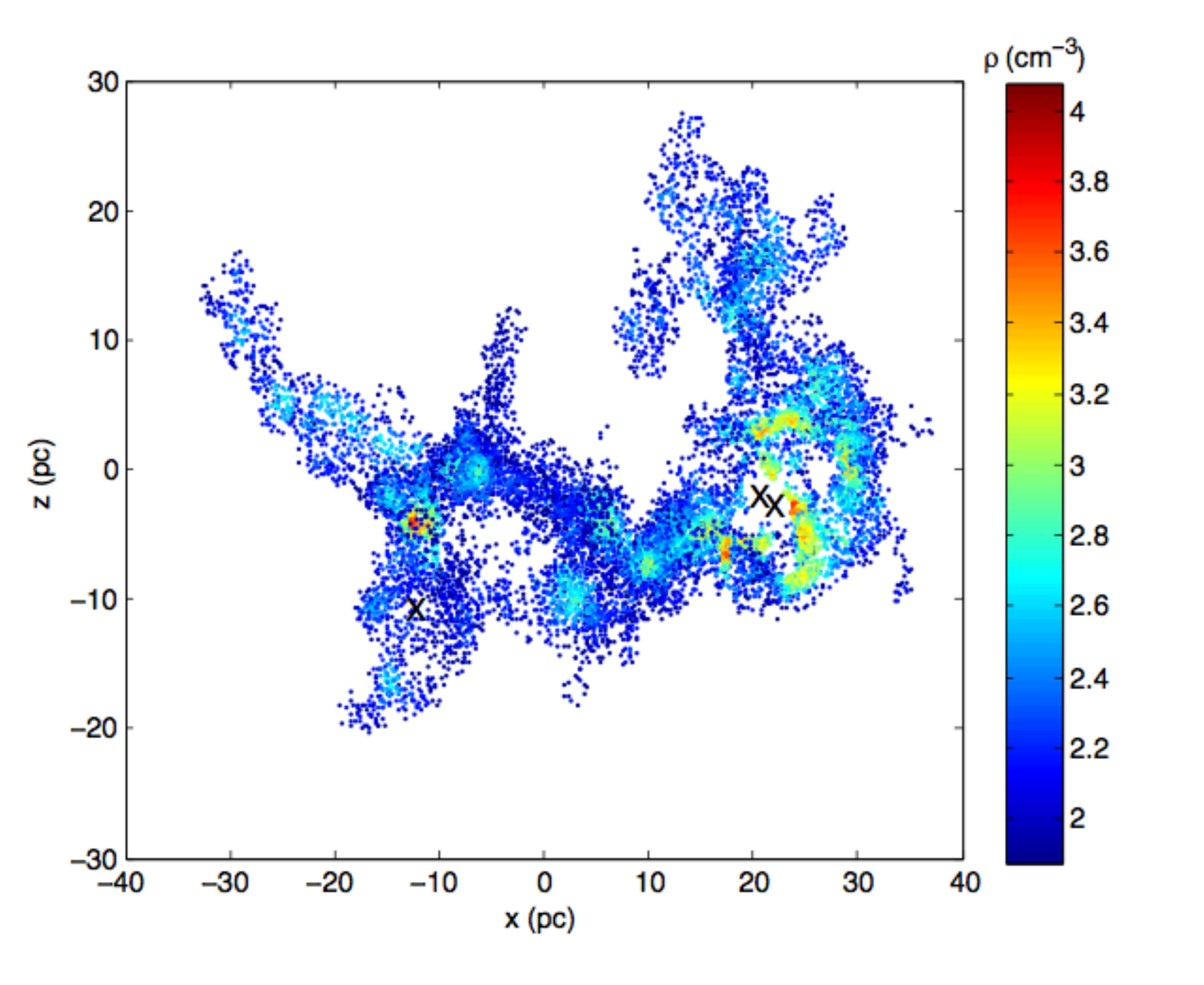}}
\caption{Velocity dispersion and density profiles are shown for 7 clouds from Run 1 ($\epsilon=0.1$), with masses between $5\times10^4-5\times10^5$ M$_{\odot}$. The dashed black line in the upper panel shows $\sigma=0.7 (r/1$pc$)^{0.5}$ km s$^{-1}$, whilst the dashed black line in the second panel shows $\rho\propto r^{-2}$. The lower panel shows one of these clouds in the plane of the disc (third panel) and in the $xz$ plane (fourth panel). The mass of this cloud is $5 \times 10^4$ M$_{\odot}$, and it is resolved with $\sim 1.3\times 10^4$ particles. The colour scale shows the density. Density peaks are situated at various positions in the cloud, some appearing to lie close to holes formed by stellar feedback (crosses, see text).}\label{internal}
\end{figure}

As well as resolving much smaller clouds, our resolution now allows us to resolve GMCs with 10,000's, even 100,000's of particles. With this resolution, we can now consider the internal properties of GMCs. In Fig.~\ref{internal} we show velocity dispersion and density profiles for a number of clouds selected from the $\epsilon=0.1$ model, Run 2. We use Run~2 simply because it contains a larger number of massive clouds. We also show an example of one of the clouds selected in the lower panels of Fig.~\ref{internal}. The clouds lie in the mass range $5\times10^4-10^5$ M$_{\odot}$ (resolved by $1-3\times 10^4$ particles), except for one cloud which is $4\times 10^5$ M$_{\odot}$, resolved by $10^5$ particles. In calculating the velocity dispersion and density profiles, we took the radius from the densest particle in each cloud, except for two cases where the densest region was right on the edge of the cloud. The spherically averaged profiles we show are somewhat simplistic given the asymmetric structure of the clouds, but give some indication of the overall structure, and can be compared with similar measures in other models and observations.

The velocity dispersion profiles  (top panel) on average show a trend of roughly $\sigma \propto r^{1/2}$, as expected from observations (e.g. \citealt{HB2004}). This trend is clearer compared to that for the whole population of clouds (Fig.~\ref{clouds_fback}), where there is considerable scatter. In terms of a turbulent power spectrum, the observed relation most resembles a power law of $P \propto k^{-4}$. This is commonly used in simulations of molecular clouds, although the profiles here are more irregular than would arise with a simple power law\footnote{\citet{Raposo2015} compare the evolution of molecular clouds from galaxy simulations to clouds modelled as spheres with a turbulent velocity field.}.
The shape of each individual profile is quite irregular whilst there is some scatter between the different profiles at a particular $r$. All the profiles flatten off at larger radii, likely due to the edge of the cloud being reached.

The density profiles are shown in the middle panel of Fig.~\ref{internal}. The profiles show a rough power law dependence, of $ \rho \propto r^{-\alpha}$, where $1.5<\alpha<2$. This is in fairly good agreement with other simulations and observations determining the density profile of core envelopes (e.g. \citealt{Caselli2002}) and larger clumps or filaments \citep{Schneider2013,Andre2013,Gomez2014}.  Most of the profiles tend to be slightly flatter at larger radii, where they are sampling a more uniform range of densities across the cloud. Various kinks, or changes in the slopes again reflect the complex structures.
We then examined the velocity and density profiles for the most massive clouds in the other simulations. Generally we found very similar properties to those in Fig.~\ref{internal}. For the higher efficiency simulation ($\epsilon=0.4$, Run 2) the peak velocity dispersions were slightly higher, $\sim5$ km s$^{-1}$ compared to 4 km s$^{-1}$  in the other models, as might be expected. 
Density profiles were similar in all the simulations.

We also computed radial density and velocity dispersion profiles, but using all the gas within a given radius of the centre of mass of a cloud, not just the gas selected as a cloud. In this case, the velocity dispersion profiles still increased at large radii, leading to a continuation of the $\sigma \propto r^{1/2}$ scaling relation. They were unchanged at small radii. The main difference with the density profiles is that they are slightly flatter at large radii, as would be expected from including more gas at larger radii. Eventually we would expect the density profiles to completely flatten out, as a fuller range of ISM densities is sampled.

All the clouds exhibit multiple density peaks, and holes, indicative of bursts of star formation occurring over time. An example structure of one cloud is shown in the lower panels of Fig.~\ref{internal} (one of the smaller clouds in our sample). For ease of viewing, the scale is changed to parsecs, centred on (0,0,0) pc. Three very recent feedback events associated with this cloud were identified, two in the top half of the cloud (in $xy$ space) from around 0.5 Myr ago and the other in the lower half from around 1 Myr ago. The positions are difficult to determine with complete precision, as the cloud structure changes with time, but the crosses indicate the equivalent part of the cloud where feedback occurred. Interestingly there are quite high densities surrounding a hole on the right hand side of the figure, perhaps dense gas which has either survived after feedback events, or dense gas that has been accumulated by feedback events. There is also a strong peak in the north of the cloud, which seems to be less obviously spatially correlated with stellar feedback. 

\section{Star formation rates}
\subsection{Star formation rates at different resolution, and with different feedback recipes}
\begin{figure}
\centerline{\includegraphics[scale=0.4]{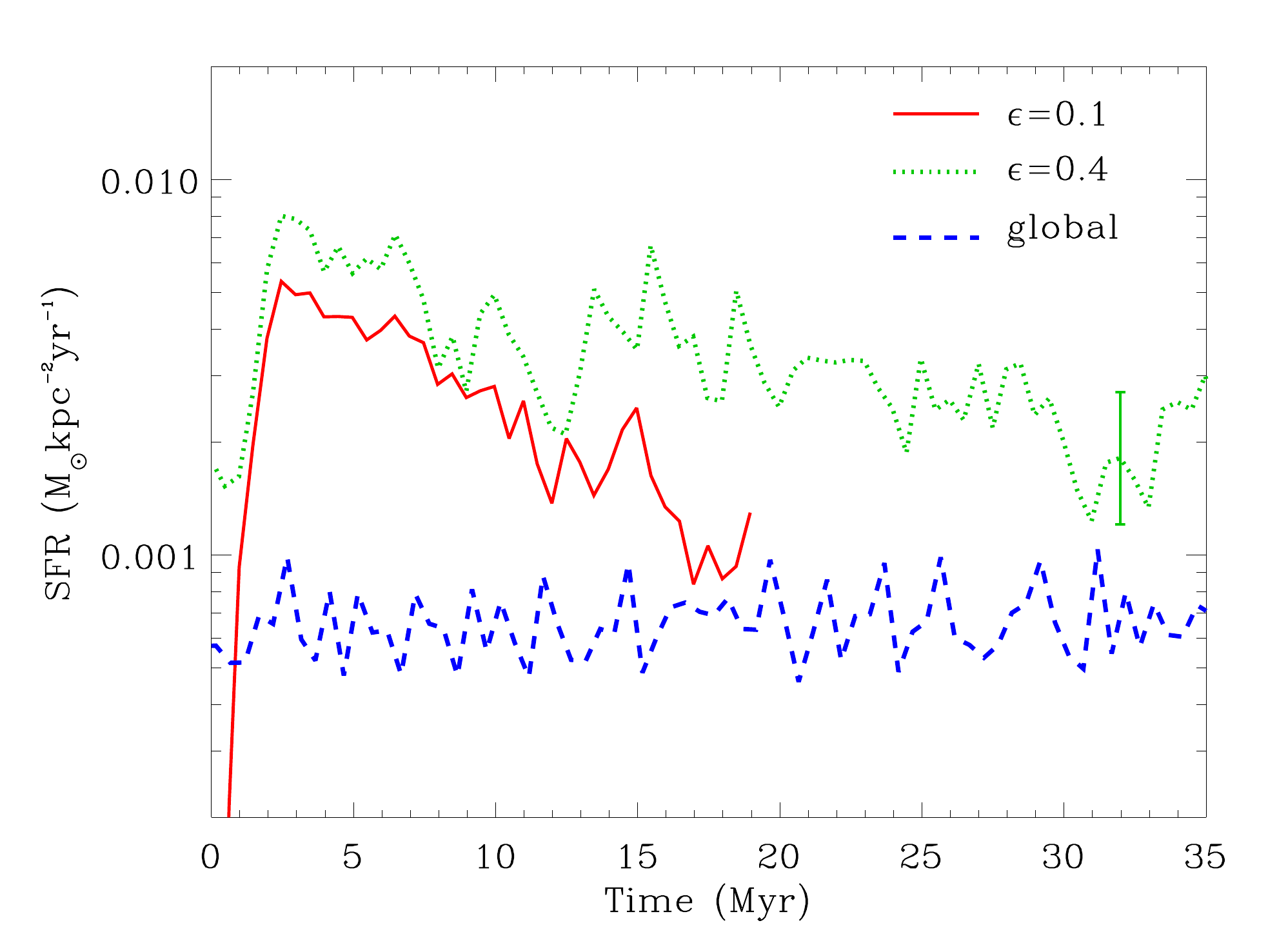}}
\centerline{\includegraphics[scale=0.4]{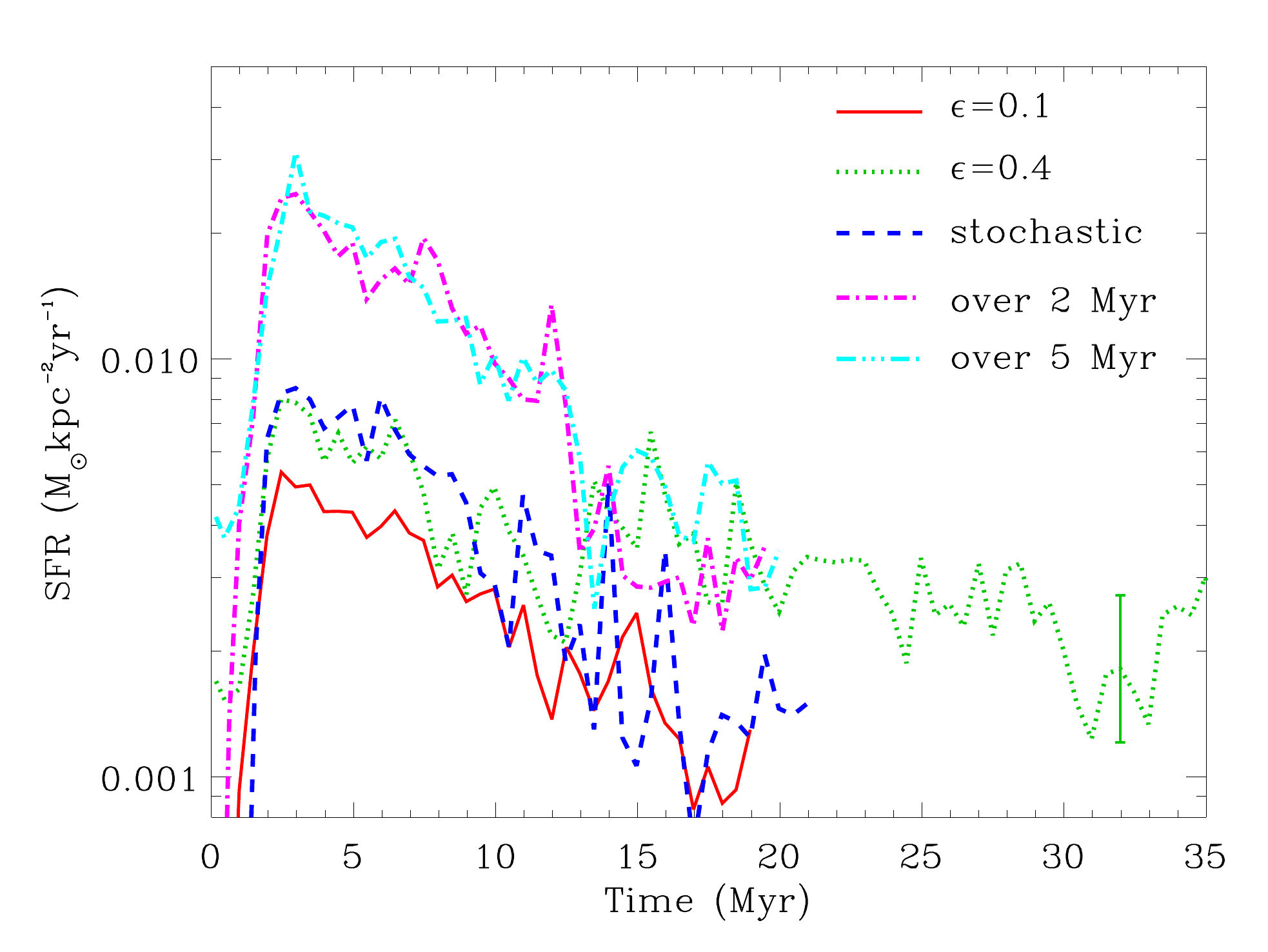}}
\caption{The star formation rates versus time are compared for the global simulation and two of the resimulations (Runs 1 and 2) in the top panel. Note that for the global simulation, the timescale actually represents from 200 to 235 Myr in the simulation. The star formation rates are compared for the models with different feedback prescriptions in the lower panel. The error bar indicates the typical 1$\sigma$ uncertainty due to the uncertainty in area, and $\epsilon$.}
\label{sfr}
\end{figure}

In this section, we study the star formation rates in the global simulations, and with the different feedback prescriptions. We compare the star formation rate from two resimulations (Runs 1 and 2 with instantanaeous feedback) with the global simulation in Fig.~\ref{sfr} (top panel). The global simulation shows a very steady star formation rate, but note that this period is over 200 Myr into the simulation, and the star formation rate is higher earlier (see \citealt{Dobbs2011new}). For both the global and small scale simulations, the star formation generally tends to peak at the beginning of simulations, and then level out, so the star formation rates in Fig.~\ref{sfr}  should be compared at later times. The star formation rate in the global simulation is slightly lower , but compares reasonably well with Run 2 which used lower feedback  at later times. Given uncertainties in the feedback scheme, and the variations with time, the star formation rates are roughly consistent between the global and resimulations.

Comparing the models with instantaneous feedback (Runs 1 and 2), the case with a higher star formation efficiency ($\epsilon=0.4$, Run 1) has a higher star formation rate compared to with $\epsilon=0.1$ (Run 2), although only by a factor of 2. This is again in agreement with \citep{Dobbs2011new}, who proposed that star formation was largely self regulating. The star formation rate for the model with stochastic feedback (Run 3, Fig.~\ref{sfr} lower panel) shows a similar amount of star formation, although the star formation rate starts higher and decreases more rapidly compared to Runs 1 and 2. The models with energy input over time have a slightly higher rate (by a factor of around 2 to 3), as again the stellar feedback seems to be less disruptive when added over time, so gives a higher star formation rate for a similar effect on the structure on the gas compared to the other simulations. However by our end time of 20 Myr, these models have star formation rates in closer agreement with the other simulations.

As well as noting that our simulations produce lower star formation rates compared to \citet{Bonnell2013} and \citet{vanLoo2013}, in \citet{Raposo2015}, we simulate individual clouds from both \citet{Dobbs2013} and this paper, without feedback. There we find star formation rates about 100 times higher than those in Fig.~\ref{sfr}, although the typical profile of the star formation rate with time (a rapid initial increase followed by a slow decline) is very similar to those shown in Fig.~\ref{sfr}. This again indicates that stellar feedback is determining the magnitude of the star formation rate. 

\begin{figure}
\centerline{\includegraphics[scale=0.27]{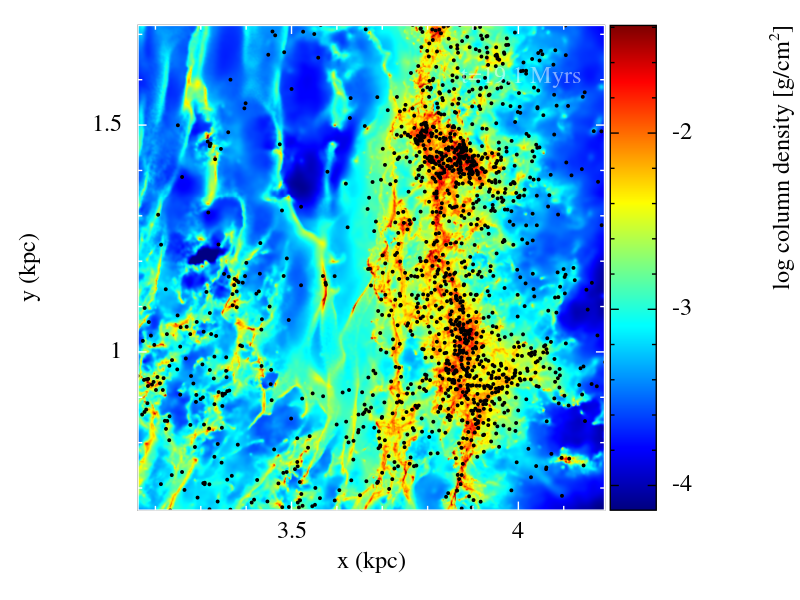}}
\centerline{\includegraphics[scale=0.4, bb=150 0 400 380]{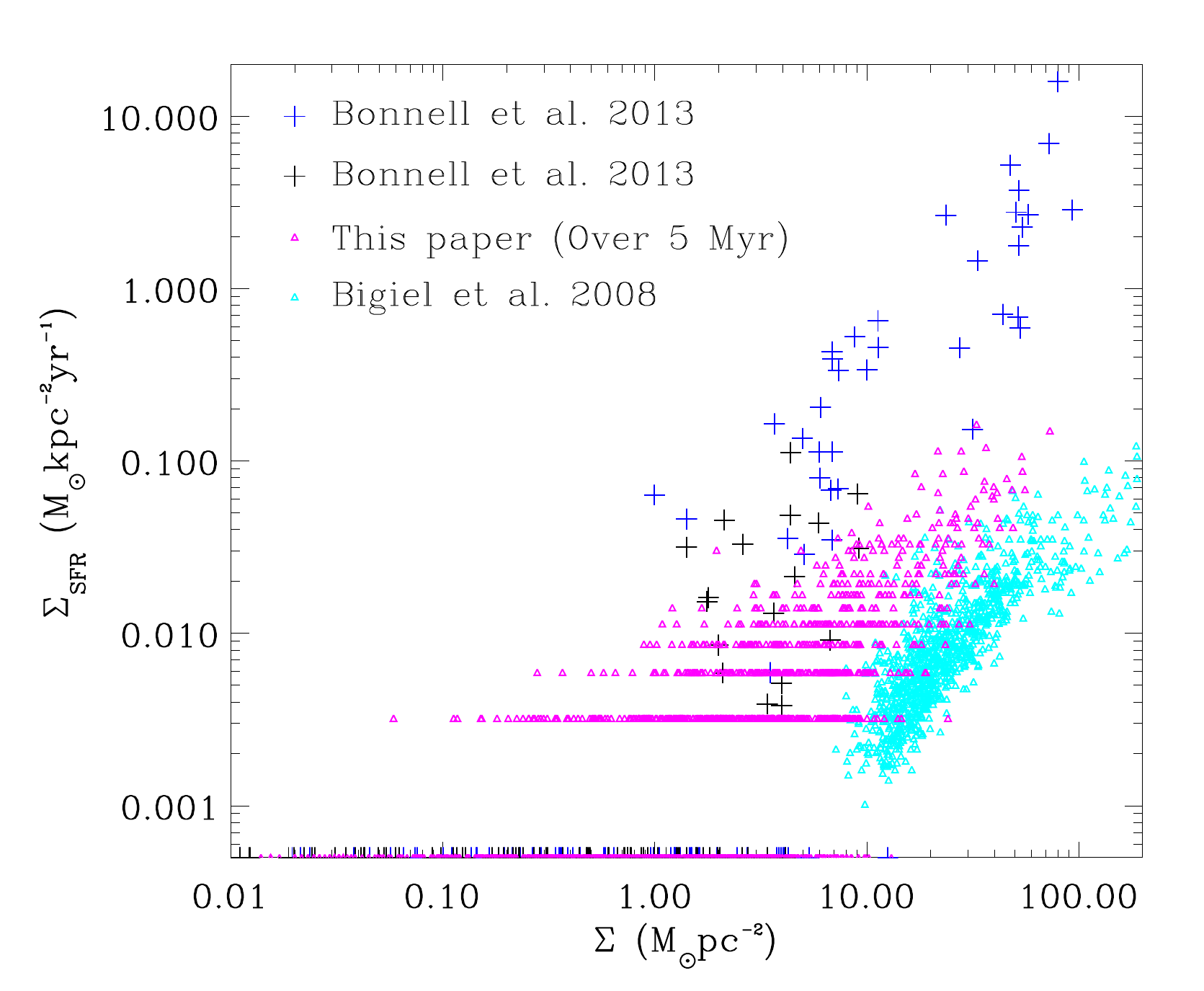}}
\centerline{\includegraphics[scale=0.4, bb=150 0 400 380]{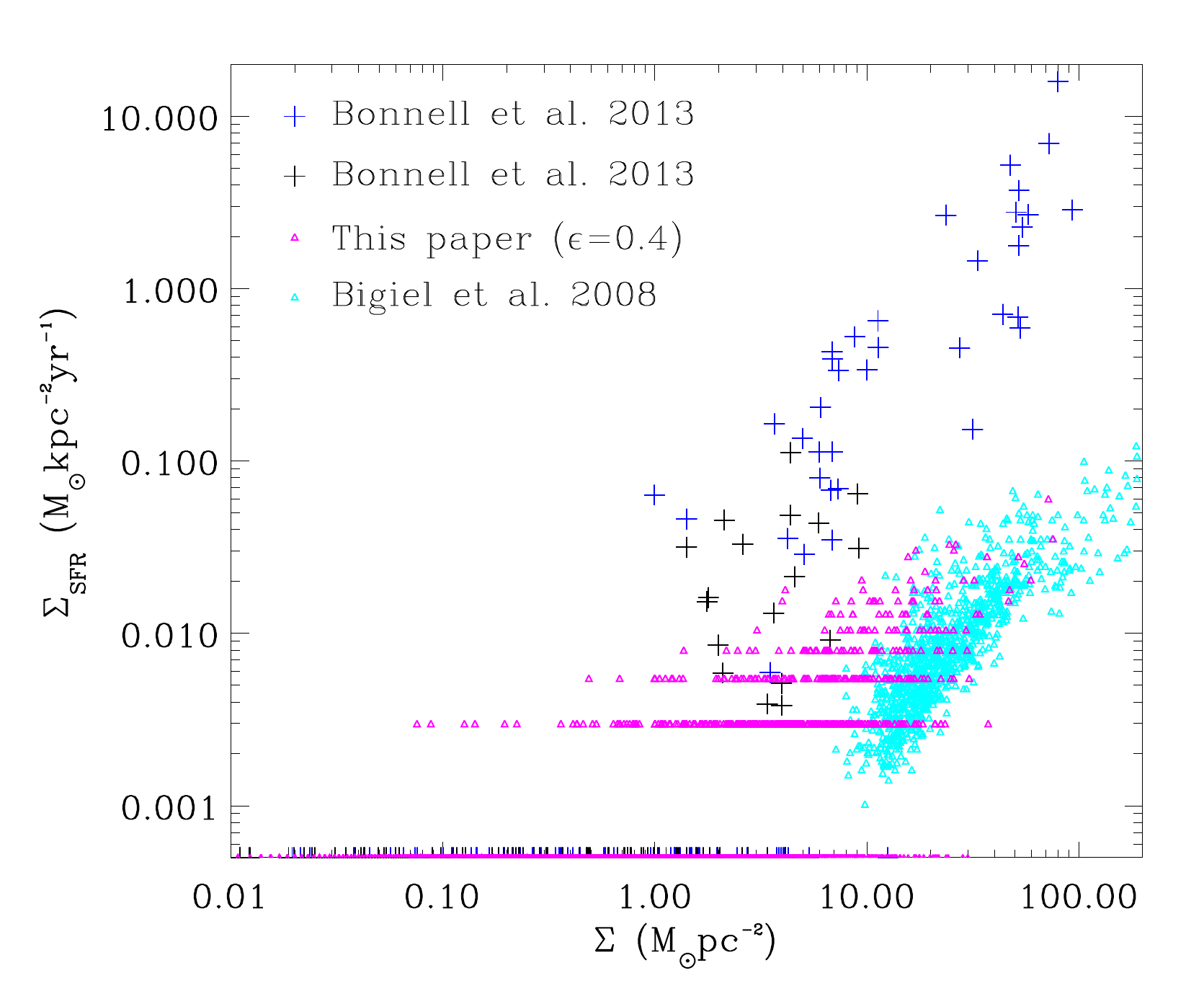}}
\caption{A column density plot is shown for Run~5 which includes star particles in the top panel (star particles are shown in black). The lower panels show the star formation rate versus (total) gas surface density fro Runs~5 and 1S. The lower panels include points from two models (with different mean surface densities) from \citet{Bonnell2013} and observational data from \citet{Bigiel2008}. The results from this paper are clearly in better agreement with \citet{Bigiel2008}. Null points for the models are shown at $\Sigma_{SFR}=5\times10^{-4}$ M$_{\odot}$kpc$^{-2}$Myr$^{-1}$. Null points for the observations are not shown. For the results in this paper, the star formation rates are set by integer multiples of star particles of fixed mass, hence the discrete sets of values.}\label{ks}
\end{figure}
\subsection{The Schmidt-Kennicutt relation}
Our simulations with higher resolution, and the inclusion of star particles, also allow us to study the variation of star formation rate with surface density (the Schmidt-Kennicutt relation) more robustly than before (see e.g. \citealt{Dobbs2011new} where we only considered global star formation rates).
In this section, we use models with star particles included. We take Run 5, where we consider star formation over a 19 Myr period, and Run~1S where we reran Run 1 (with instantaneous feedback and $\epsilon=0.4$) between 19 and 27 Myr, including star particles. Fig.~\ref{ks} (upper panel) shows a column density map of a region of Run~5, including the star particles. As expected they appear concentrated in the spiral arms, and can be seen associated with dense inter-arm structures. To produce a Schmidt-Kennicutt type figure, we divide the entire region into 50~pc cells, over which we determine the surface density, and the star formation rate. We compute the star formation rate simply by counting the number of star particles in each cell (so they will have ages up to 8 Myr for Run 1S and 19 Myr for Run 5). Our results are shown in Fig.~\ref{ks}, lower panels, with points also from models by \citet{Bonnell2013} and the observations of \citet{Bigiel2008}. The results from the resimulations, which include stellar feedback, are clearly in better agreement with the observations compared with \citet{Bonnell2013}. As discussed previously, the star formation rates are considerably lower. The star formation rates are still a little high compared to the observational data for Run~5, but are in quite good agreement for Run 1S. It is difficult to compare the shape of the distribution of the simulated points with observations, and the simulations are limited from obtaining high gas surface densities by the inclusion of feedback at such densities, but there is a tendency for distribution to curve to a shallower slope at around 10 M$_{\odot}$pc$^{-2}$ in both the observational and simulated data. Cells at very low surface densities containing star particles tend to reflect the movement of star particles away from their birth sites, or the action of feedback blowing out a hole.

We investigated star formation rates using boxes of different sizes. For a cell size of 100~pc, we simply obtain less scatter.  For a cell size of 500~pc, more comparable with the resolution of \citet{Bigiel2008}, we do not obtain such high surface density points, so it was difficult to make a comparison.
We also tried plotting the star formation rate versus molecular hydrogen surface density. This yielded points in a similar location to those shown in Fig.~\ref{ks} (lower panel), but again with a rather limited range of surface density and star formation rates to offer any insights.

\section{Conclusions}
We have investigated the interstellar medium on scales which incorporate galactic dynamics whilst resolving clouds from a few 100's~M$_{\odot}$ to $10^6$~M$_{\odot}$. We are able to model this regime by resimulating a section from a previous isolated galaxy simulation at higher resolution, achieving a particle mass of 3.85 M$_{\odot}$. We still keep a relatively simple feedback scheme, although we do investigate different feedback prescriptions and the outcome with feedback spread over different timescales. However our aim here is to study the properties of the interstellar medium with our large scale feedback (i.e. supernovae and winds) resolved very well, rather than add smaller scale processes which are poorly resolved. This also means we can directly compare the global simulation and resimulations.

We find overall good agreement with the results of the original global calculations, and our resimulations, in terms of the properties of GMCs. This suggests that the properties of our global simulations are reliable. The main exception appears to be that the velocity dispersions of poorly resolved clouds can be problematic. The simulations presented here also show that the trends in the properties of molecular clouds appear to continue down to clouds $<1000$ M$_{\odot}$. The clouds show increasing scatter of $\alpha$ with lower masses, levelling off around 1000 M$_{\odot}$, roughly constant surface densities at all masses, and a mass spectrum of $\sim M^{-2}$.

We clearly see much more structure in our resimulations than the global simulations, and we can resolve GMCs in much more detail. Tracing from a peak in density, we find density and velocity dispersion profiles of around $\rho \propto r^{-2}$ and $\sigma \propto r^{0.5}$. We can also see clear structure in the clouds, with holes from stellar feedback, dense regions adjacent to holes, and other regions of dense gas. The GMCs are highly structured, with multiple holes and dense peaks, indicating multiple episodes of star formation. Our resolution also demonstrates the ubiquity of filaments in the ISM, although we noted that filaments were at least easier to pick out in less chaotic, inter-arm regions. The filaments arise predominantly through shearing of dense regions, such as GMCs, GMAs due to the galactic rotation, but there is some evidence that feedback shapes interstellar filaments, in particular leading to a few vertically aligned filaments. Filaments likewise can show clear holes originating from stellar feedback, whilst in the vertical direction clear shells are seen, in some cases broken where gas is escaping above and below the plane of the disc.

Lastly we studied the star formation rates in the simulations, and the Schmidt Kennicutt relation. In comparison to previous simulations that study similar scales, but without feedback, we get much more realistic star formation rates, with lower star formation efficiencies. 

Our results show little dependence on the feedback scheme used. We see that feedback with a higher star formation efficiency generates more hot gas, larger bubbles, and a slightly higher velocity dispersion.  
Having the feedback spread over time is likely more realistic, but does not greatly effect our results. 
Potentially, a more dramatically different feedback scheme, which includes for example more supernovae occurring when stars are no longer associated with clouds, could induce different results.
In the simulations presented here though, the results seem most dependent on the parent galaxy model, and the properties of the gas and dynamics. We do not include smaller scale feedback processes such as photo-ionisation, but leave these for future work. We also present our results here in terms of total densities, rather than in terms of HI, H$_2$ and CO (c.f. Duarte-Cabral 2014, submitted), but again we leave this for future work.

\section{Acknowledgments}
I would like to thank the referee for a constructive report, which has helped significantly improve this paper. The calculations for this paper were performed on the DiRAC machine `Complexity', and the supercomputer at Exeter, a DiRAC jointly funded by STFC, the Large Facilities Capital Fund of BIS, and the University of Exeter. CLD acknowledges funding from the European Research Council for the FP7 ERC starting grant project LOCALSTAR. CLD thanks Ian Bonnell for providing data for Fig.~\ref{ks}, and Jim Pringle, Sarah Ragan and Ana Duarte-Cabral for useful comments and discussions. Several figures in this paper were produced using \textsc{splash} \citep{splash2007}.
\bibliographystyle{mn2e}
\bibliography{Dobbs}

\bsp
\label{lastpage}
\end{document}